# Title:

**High Performance Implementation of the Hierarchical Likelihood for Generalized Linear Mixed Models. An Application to estimate the potassium reference range in massive Electronic Health Records datasets.**

**Authors:**
Cristian G. Bologa[*], Vernon Shane Pankratz*, Mark L Unruh[*], Maria Eleni Roumelioti[*], Vallabh Shah[*,¶], Saeed Kamran Shaffi[*], Soraya Arzhan[*], John Cook[†], Christos Argyropoulos[*,‡]

*Department of Internal Medicine, University of New Mexico School of Medicine, MSC10 5550

1 University of New Mexico Albuquerque, NM 87131, USA
¶ Department of Biochemistry and Molecular Biology, University of New Mexico School of Medicine MSC08 4670 1 University of New Mexico Albuquerque, NM 87131
† Singular Value Consulting

‡ Corresponding author: cargyropoulos@salud.unm.edu

## Abstract:

**Background:** Converting electronic health record (EHR) entries to useful clinical inferences requires one to address the poor scalability of existing implementations of Generalized Linear Mixed Models (GLMM) for repeated measures. The major computational bottleneck concerns the numerical evaluation of multivariable integrals, which even for the simplest EHR analyses may involve millions of dimensions (one for each patient). The hierarchical likelihood (h-lik) approach to GLMMs is a methodologically rigorous framework for the estimation of GLMMs that is based on the Laplace Approximation (LA), which replaces integration with numerical optimization, and thus scales very well with dimensionality.

**Methods:** We present a high-performance, direct implementation of the h-lik for GLMMs in the R package TMB. Using this approach, we examined the relation of repeated serum potassium measurements and survival in the Cerner Real World Data (CRWD) EHR database. Analyzing this data requires the evaluation of an integral in over 3 million dimensions, putting this problem beyond the reach of conventional approaches. We also assessed the scalability and accuracy of LA in smaller samples of 1 and 10% size of the full dataset that were analyzed via the a) original, interconnected Generalized Linear Models (iGLM), approach to h-lik, b) Adaptive Gaussian Hermite (AGH) and c) the gold standard for multivariate integration Markov Chain Monte Carlo (MCMC).

**Results:** Random effects estimates generated by the LA were within 10% of the values obtained by the iGLMs, AGH and MCMC techniques. The H-lik approach was 4-30 times faster than AGH and nearly 800 times faster than MCMC. The major clinical inferences in this problem are the establishment of the non-linear relationship between the potassium level and the risk of mortality, as well as estimates of the individual and health care facility sources of variations for mortality risk in CRWD.

**Conclusions:** We found that the direct implementation of the h-lik offers a computationally efficient, numerically accurate approach for the analysis of extremely large, real world repeated measures data via the *h-lik* approach to GLMMs. The clinical inference from our analysis may guide choices of treatment thresholds for treating potassium disorders in the clinic.

# Background:

Electronic Health Records (EHR) have been adopted near universally in medical practices across the United States. They contain information about vital signs, lab results, medical procedures, diagnoses, medications, admissions, healthcare facility features (e.g., type, location and practice) and individual subject outcomes (e.g., death or hospitalizations). Due to the large number of patients, and the equally large number of features collected in each subject, EHR "big data" is increasingly being mined hoping to gain useful clinical insights that enhance patient safety, improve health care quality and even manage costs [1–4]. A unique feature of EHR data that complicates analytics is the repeated measures and unbalanced nature of the data, featuring multiple and unequal number of observations in individual patients (IP), seen by different healthcare practitioners (HCPs) across healthcare facilities (HCFs). However existing approaches to big data such as deep learning[5] overlook this feature of EHR data, missing on a unique opportunity to mine variation at the level of IP/HCPs or HCFs. Subject or healthcare facility focused inference requires the deployment of Generalized Linear Mixed Models (GLMM) for the analyses of repeated measures at the group (IP/HCP/HCF) level. GLMMs can tackle the entire spectrum of questions that a clinical researcher would want to examine using EHR data. Such questions typically involve the analyses of continuous (e.g., biomarker values, vital signs), discrete (e.g., development of specific diagnoses), time to event (e.g., survival) or joint biomarker-discrete/time-event data. Addressing these questions with GLMMs necessitates the evaluation of high-dimensional integrals, which even for the simplest EHR analyses may involve millions of dimensions (e.g., one for each patient). The implementation of GLMMs in existing statistical environments (e.g. SAS, R) has been shown to scale poorly, when repeated measures for more than a few thousand individuals are analyzed [6, 7]; due to the "curse of dimensionality" numerical integration is no longer tractable either analytically or numerically in these high dimensional spaces. This lack of scalability presents a major barrier in the application of formal statistical methods to big datasets. Approximate methods for fitting large GLMMs within the computational constraints of standard multicore workstations [6–8] or even parallel architectures (e.g. LinkedIn's GLMix[9]) have been proposed in the literature. Most commonly analyses rely on specific features of the datasets to speed up calculations through data partitioning and meta-analytic techniques[6]. Other proposals revisit older approaches to statistical inference by applying the method of moments during estimation [7], fitting the models through summary statistics[8] and revisiting unsophisticated but rather robust optimization methods such as coordinate block relaxation and the more recent reincarnation of stochastic gradient descent. Empirical analyses [7] demonstrates that such approaches invariably trade statistical efficiency for speed, effectively discarding valuable information hidden in big data. Finally, quantification of uncertainty about model estimates becomes extremely challenging theoretically and, in some cases, it is not addressed at all in publications (as in GLMix).

The ideal approach to the analyses of EHR big data with GLMMs, would seek to eliminate or reduce the need for statistically inefficient approximations, while retaining the rigor, numerical precision, and uncertainty quantification measures (e.g., standard errors) that one has come to expect and trust from analyses of small to medium sized data. In this work we show how these ideals can be retained in big data analytics, by deploying theoretically rigorous estimation methods for GLMMs in a computationally efficient manner. The theoretical background for achieving these goals is based on the *Hierarchical likelihood* (*h-lik*) approach to GLMM estimation[10–12]. The *h-lik* was initially introduced as a theoretical framework for understanding the statistical properties of GLMMs, but later received attention for numerical work. In the context of *h-lik*, the Laplace approximation (LA) is used to replace multivariate integration with optimization, paving the road for computationally efficient implementations that are uniquely suited to tackle the challenges posed by EHR big data, yet to our knowledge this approach has not been applied in such a context.

**Motivating example**

The analyses reported in this paper were motivated by the following clinical considerations from the physician author's subspecialty (nephrology):

- Abnormalities in the serum potassium levels ($K^+$, dyskalemia), i.e., either low (hypokalemia) or high (hyperkalemia) are commonly encountered in clinical practice[13]. Aberrations in serum potassium will in turn interfere with the electrical conduction and cell membrane function in many tissues, including muscle, bowel and heart leading to muscle weakness/small bowel ileus, cardiac arrhythmias and even death.
- The precise relationship of $K^+$ with mortality remains poorly defined. The largest to date repeated measures study in the US examined a cohort of 55,266 individuals from a single Managed Care organization in California[14]. It demonstrated a "U" shaped curve suggesting that mortality was higher for both hypokalemia and hyperkalemia. However, this analysis was limited in examining outcomes from a single health care organization and discretized the $K^+$ measurements prior to analysis. Therefore, a continuous risk relationship that generalizes across the entire United States is difficult to infer from this publication.
- This lack of precision limits rational use of both potassium replacement in patients with low $K^+$ (hypokalemia) and potassium binding drugs in those with elevated $K^+$ (hyperkalemia). Existing and emerging therapies for hypertension[15], congestive heart failure[16–18] and therapies for diabetic kidney disease[19, 20], require the use of drugs e.g. the mineralocorticoid receptor antagonists that will variably elevate the potassium level[21]. By leveraging the EHR we hope to inform future practice guidelines for the management of potassium disorders in clinical practice. For example, these guidelines could link initiation of therapies to treat dyskalemias to the prevailing $K^+$ level, using the risk relation identified in real world data to specify treatment thresholds.

**Contributions**

These can be summarized as follows:

1) We provide a roadmap for the direct, accurate and scalable implementation of GLMMs using the *h-lik* approach. This implementation effectively maps the core concepts of the statistical theory behind *h-lik*, to calls to software libraries that approximate marginal likelihoods via the Laplace Approximation (LA). These libraries employ state of the art methods for Algorithmic Differentiation (AD), that facilitate the fast calculation of measures of uncertainty (covariance matrices/standard errors). Rather than having to worry about manually encoding a high-performance implementation, the analyst can leave the numerical subtleties to be decided by the library. This implementation is scalable enough to be deployed in standard multicore workstations available to most clinical epidemiologists and as we show below yields answers to clinically meaningful questions in an acceptable timescale. Using simulations and real world big data, we show that the results obtained by our implementation are very similar to those obtained via theoretically more accurate techniques for GLMM estimation, i.e. Adaptive Gaussian Hermite (AGH) or Markov Chain Monte Carlo (MCMC) or even the original interconnected Gamma Model (iGLM) implementation of the *h-lik* [12].
2) We address the gap in the clinical knowledge about the relation between $K^+$ with mortality adjusting for the effect of kidney function on this relation.
3) We quantify the interindividual and healthcare facility variation around the curve that relates the $K^+$ level to survival Cerner Healthfacts Electronic Health Record (EHR) database (now referred as Cerner Real World Data and abbreviated as "CRWD" in this work). Cerner is one of the major EHR used in the United States. It contains laboratory, clinical and outcomes data from nearly a third of US healthcare facilities over a period of more than 10 years. Assessment of such variability is a major advantage of GLMMs over other analytical approaches, e.g., deep learning and has implications for both policy and clinical practice.

This paper is organized as follows. In the methods section we introduce CRWD and the dataset used for our analyses. We will also also review *h-lik* for GLMMs, and alternative methodologies for estimating GLMMs (AGH and MCMC), the original (iGLM) approach for fitting the *h-lik* and describe our own implementation of the *h-lik* calculations in the R package TMB.  In the results section we compare the results obtained by TMB and the referent GLMM implementation in R using iGLMs, AGH and MCMC. Unlike the direct implementation of the *h-lik* we consider here, these three alternatives have difficulties in running to conclusion within a reasonable time scale when used in big datasets. Therefore, we used simulated datasets and small random samples (1% and 10%) of the complete dataset to compare our method against these competing approaches.  After contrasting these alternatives, we will then present our analyses of the clinical scenario focusing on the relative risk of an abnormal $K^+$ level while probing the magnitude of individual and facility sources of variation in CRWD.  We conclude by considering the impact of the proposed computational methodology for EHR big data, applications outside the narrow field of EHR analytics, and the path towards fully parallel implementations of the *h-lik* approach for GLMMs.  Parts of this work have been presented in preprint [22]  and  in abstract/poster form presented during the American Society of Nephrology 2018 meeting [23].

## Methods
**Cerner Real World Data and a "Healthy" Potassium Level.**
Cerner is a comprehensive source of de-identified, real-world data that is collected as a by-product of patient care from over 700 healthcare facilities across the United States. The relational database contains clinical records with time-stamped information on pharmacy, laboratory, admission, and billing data for over 69 million unique patients. Types of data available include demographics, encounters, diagnoses, procedures, lab results, medication orders, medication administration, vital signs, microbiology, surgical cases, other clinical observations, and health systems attributes. Detailed pharmacy, laboratory, billing, and registration data go back as far as 2000 and contain more than 630 million pharmacy orders for nearly 3,500 drugs by generic name and brand. The two largest tables in the database contain more than 4.3 billion laboratory results and 5.6 billion clinical events linked to more than 460 million patient visits. This is a rich source of repeated measures data for individual level clinical variables and outcomes. Such outcomes may include *counts* of clinical encounters (e.g. hospitalizations, emergency department visits, outpatient clinic appointments), complications of treatment and deaths/ For our analyses, we used the entire CRWD from inception until September $1^{st}$, 2016  and extracted the first $K^+$  measurement that had been obtained within 24 hours of a clinical encounter.  We restricted attention to adult (older than 18-year-old) patients who were not receiving chronic dialysis (as the latter patients often have abnormal $K^+$ values) and had received at least two $K^+$ measurements in two clinical encounters. The latter provided a bona fide repeated measures dataset to analyze with the proposed GLMM implementation.   Individuals were observed until the resolution of the clinical encounter for a total of 29,787,791 days (~81,610 patient years). After excluding missing data, this dataset included 9,935,812 observations in 3,123,457 individuals and 327 facilities. During the observation period there were 48,578 deaths, corresponding to an event rate of 59.5/100 patient years. The median/interquartile range number of repeated observations per patient were 2/1 (mean and standard deviations were 3.18/2.72) and only 25% of patients had three or more repeated measurements. Most patients in our analyses had observations from a single HCF (2,886,375/3,123,457 or 92.4%). Hence, even though most patients were "nested" inside HCF, those who had "crossed" to multiple facilities create computational bottlenecks in the analyses when random effects at both the IP and the HCF facility are considered as we discuss below.

**Generalized Linear Mixed Models for repeated measures data.**

Let $y_i = \{y_j\}_i$, $i = 1, \ldots, n, j = 1, \ldots, N_i$ denote a vector of $N_i$ repeated measures outcomes for the $i^{th}$ individual, $y_{i,j}$ the $i^{th}$ outcome from that individual and $Y$ the "stacked" outcomes for all individuals in the dataset. Furthermore, $x_i = \{x_{k,j} | x_{k_1,j} = x_{k_2,j}\}_i$, $k = 1, \ldots, N_i$ $j = 1, \ldots, p$ the *design* matrix of *fixed effects* and $z_i = \{z_{k,j}\}_i$, $k = 1, \ldots, N_i$, $j = 1, \ldots, q$ the design matrix of the random effects. In the simplest case, the later indicates the group membership of the data, e.g., which observations came from the same individual, but more complex scenarios are possible dependent on the clustering structure of the data. The $x_i$ matrix is associated with a global ("*fixed effects*") $p$ dimensional coefficient vector $\beta$ while the $z_i$ with a $q$ dimensional vector $u$ of random effects. Stacking the individual fixed and random effects matrices, leads to the matrices $X, Z$. A GLMM is defined by the following properties:

1. The expectation ($\mu$) of the data ($Y$) conditional on the random effects is given by the equation: $g(\mu) = X\beta + Zu$. The function $g(\cdot)$ is the *link* function and is determined by the nature of the regression e.g., it is the logistic map for binary classification, or the logarithmic function when modeling count data. The variance of the data is determined as a product of a *dispersion parameter* (conventionally indicated as $\phi$) and a variance function that is strictly a function of the conditional mean.
2. The probability density/mass function ($f(y_{i,j}|x_i, z_i, \beta, u, \phi)$), of the data conditional on the random and fixed effects design matrices and vectors, is a member of the exponential family. Since regression is always conditional on the design matrices, we will abbreviate this density as $f(y_{i,j}|x_i, z_i, \beta, u, \phi) \equiv f(y_{i,j}|, \beta, u, \phi)$. The nature of the data determines the exponential family, e.g. for binary classification this would be the binomial distribution, for continuous data the Gaussian distribution, and for the analysis of time-to-event or counts of events (as in the present case) the Poisson distribution[24].
3. The random effects vector is assumed to have a $q$ dimensional multivariate normal distribution with zero mean and covariance matrix $G = G(\gamma)$ where $\gamma$ is a r dimensional vector of *variance components*, i.e., $f(u|\gamma) \sim MVN(0, G)$. We stress that the specification of random effects in the *h-lik* is not restricted to the normal distribution, yet this choice is made in the mixed model implementations and we will adopt this assumption.

The GLMM can be viewed as a stochastic data generation model, in which one first samples the random effects from their multivariate (normal) distribution and then conditional on those one samples the outcome variables. The associated joint, extended or hierarchical, likelihood (see pages 101-102 [12]), is given by:

$$L(\beta, u, \gamma, \phi; Y, u) = \prod_{i=1}^{n}\prod_{j=1}^{N_i} f(y_{i,j}|\beta, u, \phi) f(u|\gamma) = f(Y|\beta, u, \phi)f(u|\gamma) \qquad (1)$$

The logarithm of $L(\beta, u, \gamma, \phi; Y, u)$, $h \equiv h(\beta, u, \gamma, \phi; Y, u) = \log f(Y|\beta, u, \phi) + \log f(u|\gamma)$ is the *hierarchical (log)likelihood function* and plays a key role in calculations.

**GLMMs inference requires high-dimensional integrations.**

GLMM inference corresponds to estimating the values of the variance components ($\gamma$), dispersion parameter(s) ($\phi$) the fixed ($\beta$) and the random effects ($u$). *Maximum likelihood estimation* (MLE) for the variance components and the fixed effects involves maximization of the marginal likelihood $L(\beta, \gamma, \phi; Y)$:

$$\max_{\beta,\gamma,\phi} L(\beta, \gamma, \phi; Y) = \max_{\beta,\gamma,\phi} \int \exp(\log f(Y|\beta, u, \phi) + \log f(u|\gamma))\, du \qquad (2)$$

The integral in (1) is typically over the high dimensional space of the random effects and can only be approximated numerically. In particular, evaluation of such integrals in closed form are typically available for only a few combinations of random effects and response distributions. *Quadrature* approximations replace integrals via *weighted* sums over predefined abscissas ("nodes") that are placed over the domain of integration. The quality of the approximation and the computational resources required to evaluate the integral will scale up with the number of nodes (order) used by the quadrature rule. Integration of GLMMs in most statistical computing environments is based on *Adaptive Gauss-Hermite* quadrature[25–28]. AGH carefully *centers* the location of the nodes by finding the values of the random effects (conditional modes, $\hat{u}$) that maximize the hierarchical log-likelihood function and *scales* them according to the curvature of that function around its mode. These scaling factors are determined from the second derivatives (entries of the Hessian matrix) of the hierarchical loglikelihood function around its maximum. While a significant advance in numerical mathematics for GLMMs, the combination of MLE and AGH is not exactly problem free:

a. AGH is still as subject to the curse of dimensionality as the non-adaptive versions of Gaussian quadrature: it provides a more efficient way to spend one's fixed computational budget, by focusing on the area of the integrand that is substantially different from zero. For all intents and purposes, estimation of GLMMs with large number of random effects will be limited to a single node (AGH1) which is just the Laplace Approximation see ([29–31] and below)
b. The variance components will often be estimated with considerable bias, especially for non-Gaussian outcomes (e.g. logistic or Poisson) models and few observations per random effect [32, 33]. The Restricted Maximum Likelihood (REML) approach introduced in the next section [34, 35] may be used to reduce bias.
c. The curse-of-dimensionality often makes the use of higher order (more accurate) AGH approximations impractical: applying an $m$ order rule in $q$ dimensions will require $m^q$ evaluations of the integrand thus quickly exhausting computational resources. Higher order AGH rules are only practical for GLMMs with *nested* random effects: in this case (1) assumes the special form of nested univariate integrations and numerical integration requires $m^2$ evaluations of the joint likelihood, irrespective of the dimension of the random effects.

**H-likelihood inference for GLMMs**

The *h-lik* is an inferential procedure for GLMMs, in which the relevant calculations proceed in stages[10, 10, 11, 35, 36]. *h-lik* computations are based on the class of functions $p_w(h(\theta, w)) = p_w(h)$ defined as:

$$p_w(h) \equiv \left[ h - \frac{1}{2}\log|-H(h, w)/2\pi| \right]_{w = \widehat{w}_\theta} \tag{3}$$

where $|\cdot|$ is the determinant, $H(h, w)$ is the Hessian (matrix of second derivatives) of the function $h$ with respect to the argument $w$. The expression in (3), when evaluated at the point $\widehat{w}_\theta$ which verifies the (non-linear) system of the score equations $\partial l(h)/\partial w = 0$ for a given $\theta$, is an adjusted profile log-likelihood [37] that allows the elimination of the nuisance effects $w$, through marginalization (random effects) or conditioning (fixed effects)[11]. These eliminations are required during steps 1-2 of the *h-lik* inferential procedure:

1. Inference about the variance and dispersion components is based on maximizing $p_{\beta,u}(h(\beta, u, \gamma, \phi; Y, u)) = p_{\beta,u}(h)$ with respect to $\gamma, \phi$, yielding the corresponding REML point estimates $\hat{\gamma}, \hat{\phi}$. The uncertainty (standard errors) of these estimates is obtained from the Hessian of $p_{\beta,u}(h)$ at the optimum.

2. Inference about the fixed effects is based on maximization of the marginal likelihood (1). This maximization can be approximated via the maximization of $p_u\left(h(\boldsymbol{\beta}, \boldsymbol{u}, \hat{\boldsymbol{\gamma}}, \hat{\phi}; \boldsymbol{Y}, \boldsymbol{u})\right) = p_u(h_{\hat{\gamma}, \hat{\phi}})$ over $\boldsymbol{\beta}$ by using the plugging in the (point) estimates $\hat{\boldsymbol{\gamma}}, \hat{\phi}$. Standard errors for the fixed effects at the optimum $\hat{\boldsymbol{\beta}}$ are computed through the Hessian of $p_u\left(h(\boldsymbol{\beta}, \boldsymbol{u}, \hat{\boldsymbol{\gamma}}, \hat{\phi}; \boldsymbol{Y}, \boldsymbol{u})\right)$.
3. The random effects are obtained by optimizing the hierarchical log-likelihood $h(\hat{\boldsymbol{\beta}}, \boldsymbol{u}, \hat{\boldsymbol{\gamma}}, \hat{\phi}; \boldsymbol{Y}, \boldsymbol{u})$ given the estimates of the variance/dispersion parameters and the fixed effects [10].

Alternatively, one may *jointly* optimize the hierarchical likelihood, $h(\boldsymbol{\beta}, \boldsymbol{u}, \hat{\boldsymbol{\gamma}}, \hat{\phi}; \boldsymbol{Y}, \boldsymbol{u})$, for $\boldsymbol{\beta}, \boldsymbol{u}$ and thus obtain estimates for fixed and random effects in one pass[10]. Despite its computational attractiveness, this joint optimization may lead to non-negligible data for sparse, binary datasets, with small number of repeated observations per cluster as discussed in [12] (Chapter 6). Consequently, the appropriateness of joint maximization should be judged on a model-by-model basis.

The relation of the *h-lik* approach to the Laplace Approximation for multivariate integration is immediate. For a general q dimensional integral, the LA for a function in the variables $\boldsymbol{\theta}$, $I(\boldsymbol{\theta})$, defined via the integral of an exponentiated kernel $h(\boldsymbol{\theta}, \boldsymbol{w})$ over the variables $\boldsymbol{w}$, is given by:

$$I(\boldsymbol{\theta}) = \int exp(h(\boldsymbol{\theta}, \boldsymbol{w})) \, d\boldsymbol{w} \approx (2\pi)^{q/2} \exp(h(\boldsymbol{\theta}, \hat{\boldsymbol{w}}_\theta)) |-H(h, \hat{\boldsymbol{w}}_\theta)|^{-1/2} = \exp\left(p_w(h(\boldsymbol{\theta}, \boldsymbol{w}))\right) \quad (4)$$

Steps 1-2 of the *h-lik* approach approximate marginal likelihoods of the form $I(\boldsymbol{\theta})$ as multivariate normal distributions by finding the mode $(\hat{\boldsymbol{\theta}})$ and the associated curvature (Hessian matrix). *h-lik* further assumes that the *slice* of the joint log-likelihood $h(\boldsymbol{\theta} = \boldsymbol{\theta}_k, \boldsymbol{w})$ at *any given* $\boldsymbol{\theta} = \boldsymbol{\theta}_k$ can be approximated by a multivariate quadratic polynomial in $\boldsymbol{w}$. Under this quadratic approximation, the LA in (4) holds for all $\boldsymbol{\theta}_k$, not just for the value $\boldsymbol{\theta} = \hat{\boldsymbol{\theta}}$ that maximizes $I(\boldsymbol{\theta})$. If a sequence of $\boldsymbol{\theta}_k, k = 0, \ldots, k_{max}$ is generated by an iterative optimization procedure for $p_w(h(\boldsymbol{\theta}, \boldsymbol{w}))$, that starts from an initial value $\boldsymbol{\theta}_0$ and continues to update $\boldsymbol{\theta}_k$ until the value of $I(\boldsymbol{\theta}_k)$ no longer changes (or the maximum number of iterations, $k_{max}$, has been exceeded) then one obtains a direct implementation of the *h-lik*. The optimization of $I(\boldsymbol{\theta})$ takes a standard nested optimization form: during the k[th] (outer) iteration one first fixes the value of $\boldsymbol{\theta}$ at $\boldsymbol{\theta}_k$ to maximize for $\boldsymbol{w}$ (inner optimization), and then updates the working estimate of $\boldsymbol{\theta}$ to $\boldsymbol{\theta}_{k+1}$ using an optimizer of one's choice. This is a profile likelihood optimization problem that can be coded in the package TMB (discussed further below). An alternative, indirect computational approach to *h-lik* is through the interconnected Generalized Linear Model (GLM) algorithm.

**The interconnected GLM algorithm for H-likelihood inference**

While the theory of the *h-lik* is presented in terms of the LA, the actual numerical computations as introduced [10] and latter extended by numerous authors [11, 12, 38] is based on an additional approximation. Specifically, an Extended Quasi-Likelihood (EQL) approach [39] is uses to model the mean and the variance of the response variable (**Y**) and the random effects (**u**). Replacing the original GLM of the response variable and the probability density random effects by an approximation based on their first two moments, would appear a poorly justified choice given the loss of accuracy and potential for bias. However, this choice also allows the analyst to use the deviance (residuals) of a GLM as data. This feature of the EQL reduces estimation of fixed, random effects, dispersion parameters and variance components to a iterative fitting algorithm of interconnected GLMs (iGLM, see Chapter 7 in [12] and the documentation of the R implementations[40, 41] ). In this iterative algorithm, one fixes values of blocks of parameters (dispersion and variance components) to fit an augmented GLM to estimate the values of fixed and random effects. The deviance residuals from this fit are used as data for a gamma GLM leading to improved estimates of the dispersion and variance components, and the whole process is iterated to convergence.

Variations of this algorithm have been proposed to correct for errors introduced by the various approximations and the dependence of the random effects on dispersion parameters [12, 32, 36]. If correctly implemented, the iGLM approach is very fast to converge, as is typical of block-coordinate optimization algorithms[42, 43], in which one keeps certain parameters fixed, while sequentially updating others and cycling through the parameter list until estimates no longer change.

Pitfalls in the high performance implementation of the iGLM have been previously described [32] : they include a) errors in the implementation of the update equations and the derivatives (gradients) for the various GLMs, b) need of sparse representations for the matrices appearing in the gradients and updating equations, c) numerical imprecisions when finite differences are used to calculate the gradients and d) the large computational burden if finite differencing is used to calculate the Hessian, from which standard errors of estimates are computed. Current implementations of the iGLM algorithms are available in R package hglm[40] and the (archived) package HGLMMM[41] as well as GenStat. To our knowledge, a scalable, direct implementation of the hierarchical likelihood estimation for GLMMs that is not based on the iGLM algorithm has not been reported. Our implementation designated as *h-lik* for the remainder of this paper is discussed below.

**A scalable, direct implementation of the *h-lik* using AD and the LA using the R package TMB.**

For a software implementation of *h-lik* to align with the corresponding statistical theory, it is only required that inferences about the variance components be based on $p_{u,\beta}(h)$ and for the fixed and random effects on $p_u(h_{\hat{\gamma},\hat{\phi}})$ and $h$ respectively. To implement *h-lik* for very large datasets, one needs ways to calculate and optimize general functions of the form $p_w(h)$ in a manner that scales favorably with problem size. Scalability is afforded by a) bypassing finite differences for gradients during numerical optimization of the adjusted profile likelihoods, b) automatic detection of the sparsity (zero elements) in the Hessian and gradient matrices used in the optimizations to speed up matrix computations and c) parallelization of calculations of the value of the log-likelihood. The availability of all three features would free the analyst to concentrate on the specification of the likelihood and let the computer organize calculations in an efficient, expedient manner:

- The benefits of AD in the context numerical optimization are well understood: the gradients used during optimization are computed with numerical accuracy that rivals that of hand-coded analytical expressions while the computational cost for obtaining these gradients is of the same order as evaluating the original function. The latter feature should be contrasted to finite differencing, in which the cost increases with the number of dimensions) [44–46] . Finally, optimized, automatically generated software code for the Hessian also becomes available to the analyst at zero cost.
- Automatic generation for the Hessian is beneficial because a) it allows the use of second order optimization methods (e.g., Newton algorithm) for the inner iteration of $p_w(h(\boldsymbol{\theta}, \boldsymbol{w}))$ optimization and b) provides standard errors for estimated parameters, by computing the curvature matrix the optimum.
- For GLMM applications, the Hessian matrix will be a sparse one (i.e., only a few entries will be non-zero). Computer science algorithms such as graph coloring[47, 48] can be used to automatically detect the sparsity pattern of the Hessian *for each specific dataset* analyzed and tremendously speed up calculations and reduce memory requirements by ignoring elements that are zero during matrix multiplications.

Scalable optimization of marginal log-likelihoods approximated via a Laplace Approximation , was initially proposed in the context of MLE (rather than REML) estimation for mixed models [49, 50] and is the main use case of the R library TMB [51]. *Implementation* of any statistical model in TMB requires the writing of code in two languages: C++ and R. Using C++ macros, the user specifies the joint log-likelihood, receives parameters, random effects, data, and initial values from R.  R is used to prepare data, generate initial values, generate a computational tape (see below), invoke the optimizer, and post process the results. The standard TMB programming and workflow that an analyst must follow is:

1. Specify the *h-loglikelihood function* $h(\boldsymbol{\beta}, \boldsymbol{u}, \boldsymbol{\gamma}, \phi; \boldsymbol{Y}, \boldsymbol{u})$ in C++ and compile it into a dynamic library. This is done only once since the dynamic library may be used with different data for the same general model. Hence TMB code for GLMMs is reusable across application domains (and can even be used for MLE estimation if so desired).
2. Generate a computational graph ("tape") of the calculations required to analyze a *specific* dataset designating parameters as either fixed or random by calling the TMB package from within R. During tape generation, explicit dependencies between the parameters are identified by TMB and code is automatically generated for the evaluation of the gradient of the log-likelihood and its Hessian.
3. Call an optimizer of one's choice which repeatedly evaluates the adjusted profile likelihood and its gradient until convergence During this optimization, TMB will internally calculate the values of the parameters that are being eliminated by the profile likelihood using a second order optimization method (Newton's algorithm). Regardless of the choice of the optimizer, the latter must be able to utilize gradients (computed by AD) and fully leverage the potential of our implementation.
4. Runs the *report* function of TMB from within R to generate standard error of all estimates to quantify uncertainty upon convergence. There are various ways to generate these standard errors of variable speed and potential loss of accuracy: a) numerical (finite differencing) differentiation of the Jacobian , b) numerically invert an estimate of the Hessian, obtained in step 3 if a quasi-Newton optimizer (e.g., BFGS) was used for step 3, and c) multiplication of the Jacobian from the last point prior to convergence with its transpose.

This TMB workflow implements MLE for GLMMs and is equivalent to optimizing $p_{\boldsymbol{u}}(h(\boldsymbol{\beta}, \boldsymbol{u}, \boldsymbol{\gamma}, \phi; Y, \boldsymbol{u}))$ simultaneously with respect to $\boldsymbol{\gamma}, \phi$ and $\boldsymbol{\beta}$. However, we propose that an analyst can use TMB to carry out REML estimation and thus *h-lik* via a straightforward modification. First, we integrate over $\boldsymbol{\beta}, \boldsymbol{u}$ by designating both as "random" when generating the tape. Using this tape during optimization amounts to optimizing $p_{\boldsymbol{u},\boldsymbol{\beta}}(h)$  to obtain a point estimate (and a covariance matrix) for the variance components and the dispersion parameters (if any). Then fixing the value of the variance components and fixed parameters at their point estimates obtained by the first step, one generates the tape a second time designating only $\boldsymbol{u}$ as random. Optimizing this tape, is equivalent to optimizing $p_{\boldsymbol{u}}(h_{\widehat{\gamma},\widehat{\phi}})$ to obtain point estimates and covariance matrices for the fixed and random effects. The proposed implementation corresponds to the HL(1,1) approximation[52] in the hierarchy of *h-lik* algorithms[12, 32, 35].

 In our direct implementation of the *h-lik* in TMB, we typically used a constrained quasi-Newton method for the variance components and unconstrained quasi-Newton for fixed effects. Constraining the variance components to be greater than zero is typically required only if these parameters are close to the boundary value of zero, otherwise one may use any unconstrained optimization algorithm. Finite differencing the Jacobian appears to be the fastest approach to generate covariance matrices, and this is the approach we followed here.  In the *h-lik* theory, a third optimization conditional on the (fixed) values of the variance components and the fixed effects is required to obtain the value of the random effects. This is avoided by noting that the mode of $p_u(h_{\widehat{\gamma},\widehat{\phi}})$ with respects to $u,\beta$ will often be close to the mode of $h_{\widehat{\gamma},\widehat{\phi},\widehat{\beta}} = h(\widehat{\beta}, u, \widehat{\gamma}, \widehat{\phi}; Y, u)$, so that one can use the estimates of the random effects during the optimization of $p_u(h_{\widehat{\gamma},\widehat{\phi}})$.

Noting that the use sequential optimization of $p_{u,\beta}(h)$ and $p_u(h_{\widehat{\gamma},\widehat{\phi}})$ doubles the execution time, one may wonder whether the optimization for the fixed effects could be bypassed entirely. In this case, the estimates for the fixed and random effects are based on the final inner optimization step for $p_{u,\beta}(h)$. This approach effectively corresponds to the initial proposal by Lee and Nelder [10] to jointly optimize the conditional hierarchical log-likelihood $h(\beta, u, \widehat{\gamma}, \widehat{\phi}; Y, u)$ to obtain estimates for the fixed and random effects when the Hessian matrix of the hierarchical log-likelihood does not vary with the fixed and random effect estimates around the mode. It is known as the HL(0,1) approximation in the hierarchy of *h-lik* algorithms and appears to be currently implemented in the *R* package *glmmTMB* [53], though this is not explicitly stated.

In our analyses we used the high performance implementation of the BLAS libraries[54] (MKL® by Intel that are distributed with Microsoft R Open) as suggested by the authors of the TMB package. We also evaluated the speed up afforded by the parallel calculation of the *h-likelihood* function, a feature which is natively supported by TMB in multicore platforms through OpenMP.

**Verifying H-likelihood calculations via Bayesian Methods**
It has been previously pointed [11, 32, 55] out that the *h-lik* approach can be viewed as a *modal approximation* to a Bayesian analysis that uses non-informative, uniform priors for the fixed effects , dispersion parameters (if they are not known) and variance components. This connection allows one to check any method implementing the *h-lik* (e.g., the iGLM or our direct implementation) by comparing its computations against those obtained by MCMC integration of the corresponding Bayesian model. While MCMC may be considered the gold standard for high dimensional integration, it is a much more computationally demanding technique. Hence, while one can apply MCMC to small problems, one will wait for a very long time for MCMC to finish for datasets of the size we are considering. However, one can compare the results in smaller datasets that are representative of the big dataset. Such datasets may be obtained by simple random sampling without replacement from the final dataset that one would like to analyze. These analyses can serve as a sensitivity check for the quality of the numerical approximations used, as we illustrate in results section.

**Modeling the risk of dyskalemia in CRWD via Poisson GLMM.**
The data $y_i$ for these analyses were repeated observations of the disposition of patients during clinical encounters recorded in CRWD. For each individual patient, this vector is a string of zeros (if the patient was alive at the end of each encounter), possibly terminated via a one, if the patient had died. The vector of zero-one outcomes was modelled as a Poisson variable, by exploiting the link between Poisson models and analysis of survival time [24, 56], using the duration of each clinical encounter as an offset. Predictors that were coded as *fixed effects* included the $K^+$ level, the Charlson Comorbidity Index, a well validated instrument of the comorbidities of ang an individual, the prevailing level of kidney function (estimated Glomerular Rate, eGFR, computed by the validated CKD-Epi formula), patient's age, type of healthcare facility (e.g., hospital, nursing home, outpatient clinic), patient's age, race and gender. Natural splines were used to probe non-linear relationships between the outcome of interest, kidney function (eGFR) and the potassium level as well as age. During the comparative evaluation of methods for fitting the dataset, we run two separate analyses with different random effects specifications: one in which the grouping level was the individual (IP) and one in which the grouping level was the facility (HCF). These analyses allowed us to explore the scaling of performance of various estimation methods for GLMMs against the increasing dimensionality of the problem. The latter is largely determined by the number of random effects, since the vector of fixed effect coefficients is low dimensional, i.e., *p*=18 in all analyses. Clinical inference about the relation between the K$^+$ level and the risk of death was based on the full dataset using random effects at both the IP and HCF level. For both the Poisson and Logistic GLMM used in simulations, the dispersion parameter is known and fixed at the value of one, i.e., $p_u(h_{\hat{\gamma},\hat{\phi}}) = p_u(h_{\hat{\gamma}})$ and only the variance components (standard deviations of the random effects) must be estimated from $p_{u,\beta}(h)$.

**Simulated datasets for comparative evaluations of numerical methods for GLMMs**

To compare different approaches (AGH, iGLM, and the direct implementation of *h-lik*) for the integration of marginal likelihoods in GLMMs, we simulated data that recapitulate features of the EHR datasets that may be encountered in practice: unbalanced observations per IP, variable crossing over IP to different HCF for different number of individuals (100 and 1000) and health-care facilities (5 and 50). In these simulations we were particularly interested in how the structure of the regression matrix of the random effects ***Z***, affects the performance of the various estimation methods. Hence, we simulated scenarios of full nesting (an IP will visit one and only one HCF), partial crossing (25% of IP will visit two or more facilities) and more extensive crossing (75% of IP with visits at more than two HCF). In the simulations we also included a moderate number of fixed effects. The design matrix of the latter included constant terms (the intercept), binary covariates ("simulated gender"), continuous linear terms ("simulated Charlson score") and spline terms ("simulated age/eGFR/K$^+$ level"). The Poisson exponential family was used to simulate counts of discrete events (e.g., death or hospitalizations) that may be observed in the EHR: the choice of the simulated fixed effect associated with the intercept ensured that individuals would have an outcome of zero or one. Our simulation strategy (**Appendix**) yielded 200 datasets for each of the six Poisson scenarios of nesting and number of IP/HCF considered.

A critical question for the analyses of very big data is whether one adopt shortcuts e.g., the HL(0,1) instead of the HL(1,1) approach or foregoing of a third conditional optimization for the estimation of the random effects. Binary (logistic) GLMMs models with few repeated measures per cluster and a low dimensional fixed effect covariate vector are often the most challenging to fit with approximate methods[12, 32, 35, 52]. We thus simulated fifteen logistic regression scenarios using the same random effect structures (two random effects) and crossing structures like the Poisson scenarios, but fitting only a single fixed effect (the intercept). Six of the fifteen scenarios used the same variance component values ("less variable datasets") and intercept values as the synthetic Poisson data, while the remaining nine scenarios used higher values for the standard deviation of the IP random effect ("more variable datasets"). This approach yielded another 3000 simulated, artificial datasets as we describe in the **Appendix**.

For each of simulated dataset we extracted the point estimates of the fixed, random effects and variance components and computed measures of bias (the absolute standardized bias) and total accuracy (Mean Square Error, MSE) using standard formulas[57]. We also supplemented these simulations by randomly subsampling our CRWD full dataset to create two subsets that included 1% and 10% of patients. These datasets were used to contrast the inferences by the various methods in real world, moderate sized datasets.

**Software and R packages**

We used the R package lme4 [58] for AGH based inference (function glmer) and utilized quadrature with 0, 1, 5 and 9 nodes to explore the tradeoffs incurred between accuracy and speed for models with a single random effect. Note that the LA is formally equivalent to an AGH with a single quadrature node, so that the comparison between AGH1 and *h-lik* reflect the speedup due to the LA/AD combination. The "zero" order method AGH effectively ignores the integral for the random effects and thus its estimates will contain a variable and unknown amount of bias. However it appears to be one of the most robust methods available in glmer and thus was considered here. Analyses using the iGLM approach were conducted with the R package *hglm* [40], while the HL(0,1) approximation was coded directly in TMB or by specifying models in the R package *glmmTMB* [53]. Bayesian analyses were carried out by invoking the Poisson GLMM in the STAN programming language [59] as implemented in the R package *rstanarm* [60]. Like TMB, STAN codes the statistical model into a C++ language template, which is also compiled into a dynamic library that implements the calculations for the posterior probability kernel of the model. In this case, the kernel coincides with the h-likelihood function. STAN also utilizes AD to compute the gradient of the model that are used during MCMC with the No U Turn Sampler [61]. Therefore, the comparison against STAN allows an assessment of the accuracy-speed trade-off of MCMC against the *h-lik* computations when the latter are interpreted as modal approximations to a full Bayesian analysis. While the formal interpretation of the *h-lik* as a modal approximation to a Bayesian analysis requires uniform priors for the fixed and random effects, we found that using this prior for the model intercept induced very slow mixing, and long execution times during MCMC. Hence, we used the default, weakly informative prior, a Gaussian centered at 0.0 but with a standard deviation of 2.5 in the scale of the response for Bayesian runs.

**Hardware and timing**
Timing information was generated in a high-end consumer workstation equipped with the i7-5960x octacore Intel Processor equipped with 32 GB of DDR4 RAM clocked at 2133MHz. The base frequency of the i7-5960x is at 3GHz, but the processor was overclocked to 3.8 GHz for the timing experiments. For the analyses of the full dataset, we utilized a Dual 16-core Xeon Gold 5218 clocked at 2.3 GHz (turbo single core frequency of 3.9 GHz) equipped with 512GB of DDR4 RAM (clocked at 2666MHz) running Windows 10 Pro for Workstations. Large memory requirements did not allow us to fit the parallel version to the full dataset, even in this large memory machine. For the comparative evaluation of execution timing in the smaller datasets we ran the iGLM and the parallel TMB analyses thrice on the i7-5960x and the Xeon. To compare parallel and serial implementations of the *h-lik* in the TMB environment, we ran analyses of the 1% and 10% datasets on the i7-5960x using both parallel and serial versions of the code that incorporated the two random effects. In addition, we run all three datasets with the serial code on the dual Xeon. Analysis of variance was used to explore the relative impact of dataset size, parallel vs serial version of the code and computing platform, on execution ("wall") times. We collected timing information for all the steps of the TMB *h-lik* calculation e.g., the generation of the tape, optimization, and calculation of the Hessian at convergence, only the total wall time is reported. All analyses were run using Microsoft R Open v 3.5.3 and v4.02 and TMB package v1.7.14-1.7.15.

# Results:
**Structure and outcome sparsity patterns of the simulated datasets**

The distribution of observations per IP and the number of HCF "visited" per IP are shown in **Supplementary Figure 1, Appendix**; the median (Inter-Quartile Range, IQR) of visits per patient were 5(4) respectively. The median (IQR) number of HCF visited by each IP varied from 1(0), 1(1) and 4(3) for the nested, partly crossed and more crossed scenarios. The number of observations ("visits") and distinct IP per HCF are shown in **Supplementary Figure 2 in Appendix**. Each facility registered on average ~100 patients but this number varied widely (standard deviation between 23-27 visits or corresponding IQR between 33-41). The number of IP per clinic varied according to the degree of nesting; the median (IQR) was 20 (5.75), 28 (7.75) and 94 (28.5) for the nested, partly crossed and more crossed scenarios. **Supplementary Figure 3 in Appendix** shows boxplots of the number of events per observation in the simulated datasets. The Poisson datasets were the least sparse with most datasets having an average number of observations of fewer than 0.10 per observation, and binomial outcomes were sparser with number of events per observation less than 0.015.

**Comparison of *h-lik* , iGLM, AGH and MCMC based methods in the simulated datasets.**

**Poisson outcomes:** In **Figure 1** we contrast the performance of two AGH methods (the zero and first order as implemented in R's package lme4), the HL(0,1) method as implemented in the package glmmTMB, the iGLM implementation of HL(1,1) the proposed *h-lik* implementation and two MCMC estimates: that based on means of the posterior marginals and that based on posterior modes (MCMCmode). In general differences between any two methods were minimal (less than 10% of the standardized bias for each coefficient, and MSE for the fixed effects (**Figure 1**A) were similar, irrespective of the number of individual simulated patients in the datasets. On the other hand, estimates of variance components was sensitive to the number of IP random effects (**Figure 1**B), with more random effects associated with reduction in the MSE. Random effect estimates were virtually bias free for all methods, except from AGH1 (**Figure 1**C).

**Binary outcomes:** Having established the similar performance of various methods for GLMMs we then moved to compare the HL(0,1) and HL(1,1) implementations in the simulated sparse binomial datasets. In these datasets we encountered numerical instability in getting the iGLM to converge, so we only report on the TMB based implementations. In **Figure 2** we show the performance of the various methods in the "less variable" binary outcome dataset. The poor performance of the AGH1 implementation manifests both in the MSE and the absolute standard error of the single fixed effect (**Figure 2**A) and the two variance components (**Figure 2**B). Using the estimates of the fixed effects from the first optimization of the *h-lik* proposal ("h-lik-pub" in the figure) yields fixed effect estimates that are indistinguishable from glmmTMB. This is not surprising as these two methods are equivalent both mathematically and computationally. Turning to our attention to the random effect estimates (**Figure 2**C), we observe clear evidence for a small, but definite bias for the AGH1 method. On the other hand, estimates returned by the AGH0, glmmTMB, *h-lik*, the *h-lik-pub* and *h-lik-3* (which optimizes $h(\widehat{\boldsymbol{\beta}}, \boldsymbol{u}, \widehat{\boldsymbol{\gamma}}, \widehat{\boldsymbol{\phi}}; \boldsymbol{Y}, \boldsymbol{u})$ to estimate the random effects) are effectively identical. The "more variable" binary outcome dataset allowed us to test the *h-lik* implementations in very sparse binary datasets. The relevant analyses are shown in **Figure 3** and illustrate substantial bias and large MSE when the HL(0,1) method for is used for the estimation of low dimensional fixed effects against the *h-lik* implementation of the HL(1,1) method. This bias is observed for both HL(0,1) TMB based implementations, i.e. glmmTMB and "h-lik-pub". Estimation of variance components (**Figure 3**B) shows an interesting pattern: while the estimate of the largest variance component (that due to individual variability) is much more biased than the estimate of the smallest variance component, the MSE are identical. Similarly, the estimates of the random effects (**Figure 3**C) by glmmTMB are less biased than that of the *h-lik* method, yet the MSEs are identical. As the number of random effects increased, the bias in the *h-lik* method decreased and the random estimates returned by the two methods become identical (note the progressive alignment of the boxplots of the random effect estimates as the number of individuals increase from 100 to 10,000).

*In summary,* our simulations indicate that for non-binary GLMMs with somewhat sparse outcomes, one can proceed with the HL(0,1) method, as implemented in glmmTMB or via the first optimization of the *h-lik* approach for fixed effect estimation. However, for datasets with sparse binary outcomes and large number of random effects, the HL(1,1) method as implemented by our *h-lik* proposal, will yield the least bias and the smallest MSE and thus is to be preferred. Alternative numerical methods for the implementation of *h-likelihood* inference are associated with equivalent MSE performance for the estimates of variance components and random effects. Despite this equivalence, they seem to proportionate the total (MSE) error differently between bias and variance. Based on these observations, we retained the two-stage *h-lik* method for the analyses in CRWD.

**Estimates of *h-lik* against AGH and MCMC based methods for random intercept Poisson models in the CRWD.**

Analyses of the 1% and 10% datasets with glmer [7] generated warnings about at least one non-positive definitive Hessian matrix suggesting that the optimization algorithm got stuck close to the true global optimum. Furthermore, AGH1 failed to converge at all, when the measurements were grouped at the IP level. In contrast to previous reports [6], we could obtain a solution from glmer for the 10% dataset for very large number of random effects but only if the AGH5 or AGH9 method were used. Convergence failure of the AGH1-9 methods implied that their estimates may not reliable. Despite the theoretical potential for bias, the AGH0 approximation converged without warnings in the CRWD. The results of the direct implementation of *h-lik* with TMB, AGH0, and AGH9 are shown

**Figure 4** in as estimate (point) and 95% confidence interval for the two groupings and the two datasets. In general, the estimates of our *h-lik* implementation were numerically close to the AGH methods, suggesting that the nonconvergence reported by the latter would not affect inferences about the fixed effects. The difference among the different methods lies within the uncertainty of the estimate from each method; in fact, estimates were identical for most covariates up to two significant digits. Nevertheless, there were substantial differences among the methods when the estimate of the variance component was examined. **Table 1** shows the estimates for the direct implementation of *h-lik* in TMB, the iGLM implementation in the R package *hglm*, AGH0 and AGH9. The variance of the estimates could not be obtained for the other AGH methods and thus are not reported. Furthermore, we could not fit the iGLM to any dataset with IP random effects due to large memory requirements (i7-5960x) and numerical errors in platforms not limited by memory (Xeon).

**Table 1 : Estimates of the standard deviation of the variance components for single random effects Poisson models.**

| Dataset | Group | h-lik | iGLM | AGH0 | AGH9 |
|---|---|---|---|---|---|
| 1% | Individual | 0.306 (0.289) | | 0.253 | 0.260 |
| 10% | Individual | 0.469 (0.074) | — | 0.505 | 0.504 |
| 1% | Facility | 0.422 (0.085) | 0.395 (0.047) | 0.401 | 0.414 |
| 10% | Facility | 0.703 (0.061) | 0.669 (0.042) | 0.699 | 0.701 |

Increasing the size of the dataset results in a greater agreement among the three methods and reduction in the magnitude of uncertainty in the estimates of the proposed *h-lik* method. Estimates by the AGH5 method were identical to the AGH9 to three significant digits (not shown). Bayesian analyses may be considered the benchmark against which other methods of integration for GLMMs should be judged. However only the 1% dataset with grouping at the facility level could be fit in reasonable time (see timing below). The Bayesian point estimate (posterior mean) of the variance component was 0.426 (standard deviation 0.089), an estimate that was in rough agreement with the estimates generated by all the methods in **Table 2** but was numerically closer to the direct *h-lik* estimate of 0.422. The Bayesian model estimates (posterior mean and standard deviation) for the fixed effects against the direct *h-lik* and iGLM implementations are shown in **Table 2**. Point estimates and standard errors returned by the two *h-lik* implementations were in close numerical agreement. Many of the MCMC estimates, including the coefficients for the natural splines used to model the effects of age and eGFR were close to those obtained by the *h-lik* implementations. Notable exceptions were the coefficient for some of the racial groups and the coefficients for the natural splines of the potassium level. There are two related reasons for these differences: first, the lack of sufficient amount of information to estimate these coefficients from the data as evidenced by the large relative magnitude of the standard errors relative to the point estimates. This is turn causes the quadratic approximation to the likelihood (and thus the LA) to fail for these datasets. This is shown in **Supplementary Figure 4 in the Appendix** which graphs non-parametric kernel density estimates to the coefficients of the potassium spline, along with the AGH9 and the h-lik estimates. The h-lik, and to a less satisfactory extent the AGH9, clearly identify the mode of the marginal posterior distribution with precision. Since the corresponding Bayesian marginals are highly asymmetric, their means do not coincide with their modes and thus the *h-lik* methods which approximate the marginal mean by the mode fails. Despite the differences in the numerical values of the spline coefficients, the splines estimated by the proposed *h-lik* implementation and the MCMC method were nearly identical and had very similar pointwise confidence intervals (**Figure 5**).

**Table 2 : Fixed effect estimates, summarized by the mean(standard error) from Bayesian and h-lik (as implemented in this paper and the iGLM algorithm) analyses for the 1% dataset with grouping at the facility level.**

| Fixed Effects | Bayesian (MCMC) | h-lik | iGLM |
|---|---|---|---|
| Intercept | -6.214 (0.961) | -6.218 (0.994) | -6.168 (0.992) |
| Potassium (natural spline coefficient 1) | 8.062 (1.979) | 6.808 (2.462) | 6.805 (2.461) |
| Potassium (natural spline coefficient 2) | -38.03 (31.12) | -9.491 (40.24) | -10.00 (40.23) |
| Potassium (natural spline coefficient 3) | -75.33 (61.7) | -18.50 (79.78) | -19.51 (79.76) |
| eGFR (natural spline coefficient 1) | -0.406 (0.255) | -0.43 (0.251) | -0.435 (0.251) |
| eGFR (natural spline coefficient 2) | -1.872 (0.492) | -1.839 (0.505) | -1.839 (0.505) |
| eGFR (natural spline coefficient 3) | -1.141 (0.502) | -1.091 (0.509) | -1.089 (0.509) |
| Age (natural spline coefficient 1) | 0.305 (0.332) | 0.309 (0.329) | 0.306 (0.329) |
| Age (natural spline coefficient 2) | -0.086 (0.938) | -0.227 (0.913) | -0.227 (0.913) |
| Age (natural spline coefficient 3) | 0.924 (0.199) | 0.906 (0.198) | 0.902 (0.198) |
| Male Gender | 0.214 (0.093) | 0.214 (0.093) | 0.214 (0.093) |
| White Race | 0.155 (0.143) | 0.158 (0.146) | 0.160 (0.145) |
| Hispanic Race | -0.745 (0.853) | -0.457 (0.723) | -0.452 (0.723) |
| Native American | -0.233 (0.829) | 0.031 (0.731) | 0.040 (0.730) |
| Other Race | -0.618 (0.321) | -0.580 (0.323) | -0.571 (0.322) |
| Unknown Race | -0.347 (0.488) | -0.242 (0.472) | -0.238 (0.471) |
| Inpatient status | 2.404 (0.198) | 2.390 (0.194) | 2.397 (0.194) |
| Charlson Comorbidity Score | 0.098 (0.014) | 0.099 (0.014) | 0.099 (0.014) |

The predicted random effects for the 1% dataset grouped at the facility level obtained by all non- Bayesian methods were rather like those obtained by MCMC (

**Figure 6**), as evidenced by the high correlations. Bland Altman plots (lower diagonal panels in

**Figure 6**) indicate that the difference between *h-lik* and the other methods were within 10% of each other.

*In summary*, all methods appear to generate similar estimates for the fixed effects, rather similar estimates of the variance components, with the proposed direct *h-lik* implementation demonstrating the closest numerical agreement to MCMC.

**Timing comparisons between *h-lik* and the AGH methods for random intercept models.**

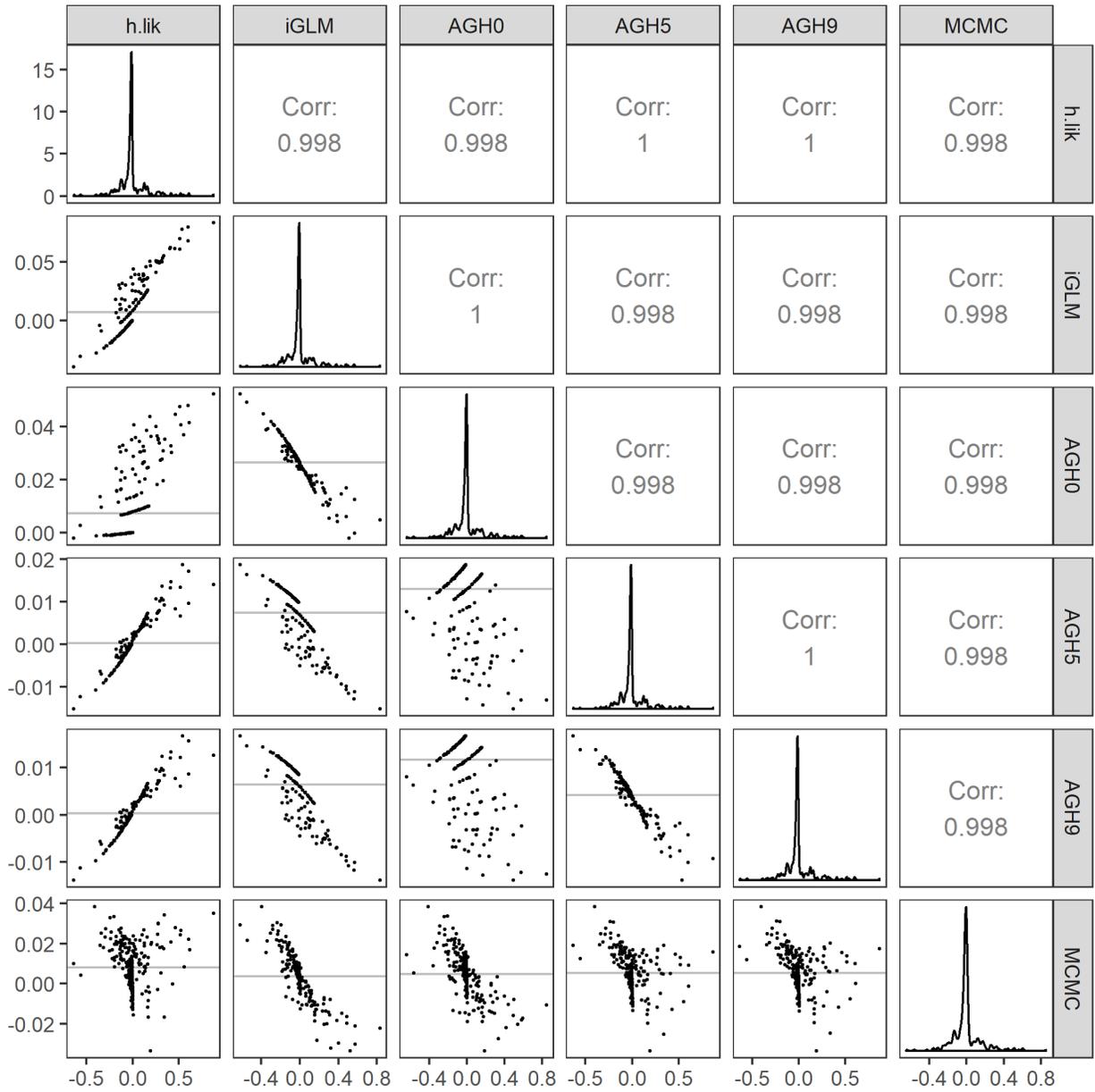

**Figure 7** contrasts the execution times between *h-lik* and the AGH methods in the i7-5960x platform. Irrespective of the size of the dataset and the grouping structure, the *h-lik* implementation was much faster than the AGH implementation in glmer. The smallest speedup was observed for the 10% dataset with grouping at the IP level (3.7x faster), but the speedup could range up to 32.5x (1% dataset, grouping at the HCF level). The direct *h-lik* implementation was only slightly slower than the iGLM in the datasets we could analyze with the iGLM.  Analysis of Variance (ANOVA) showed that the size of the dataset (F statistic 68.07) and the estimation method (F statistic 18.1) were much stronger predictors of these timing differences than the grouping structure of the data (F statistic 3.68). Examination of the timing of the three steps of the *h-lik* method showed that the greatest amount of time was spent on optimization, e.g., for the 10% dataset with random effects at the IP level, tape generation took 137sec, optimization 3327 sec and Hessian calculation/uncertainty quantification 1,844 sec. For the 1% dataset with random effects at the HCF level, the corresponding times were: 8 sec (tape generation), 17 sec (optimization), 1.7sec (uncertainty quantification). In sharp contrast, Bayesian methods were much slower than any of the other methods considered. Therefore, our analyses were limited to the 1% dataset at the HCF level. Parallel simulation of 1,000 samples from each of four independent Markov Chains took 20,517 seconds of wall time, thus was nearly three orders of magnitude slower than either of the two *h-lik* implementations.

**Potassium level, interindividual and interfacility sources of variation in the risk of death in CRWD.**

Predictors of mortality in the CRWD accounting for both IP and HCF variability by a bivariate random effects model and for the various datasets considered here are shown in **Table 2**. Increasing the dataset size, stabilized model estimates of the (log) relative risk of death associated with each of the covariates and was associated with a substantial decrease in the standard error of these estimates. There was considerable interindividual and (to a larger extent) interfacility variation in the relative risk of death, as evidenced by the magnitude of the variance components in the full CRWD dataset. The point estimate (standard error) was 0.234 (0.034) and 1.11 (0.067) for the IP and HCF components, respectively.  These estimates imply that individual factors not captured by the covariates considered in Table 1 will be associated with 2.5-fold variation in individual risk between  the 2.5% and the 97.5% risk quantiles of patients with the exact same values of these covariates. Similarly, facility level factors are an important source of variation, though in this case the relative risk of an individual evaluated in a facility at thee 97.5% quantile vs. a facility in the 2.5% quantile is 77.5-fold higher.

The coefficients of the spline associated with the $K^+$ level appear rather different among the three datasets. Nevertheless, the predicted relative risk of death for a given $K^+$ level was essentially identical between the 10% and the 100% datasets (**Figure 8**). Utilizing the full (100%) dataset resulted in a substantial reduction in the uncertainty (95% pointwise confidence interval) for a given $K^+$ level. The relative risk associated was numerically close to one for a  broad range of $K^+$ between 3.3 to 4.8 meq/l, i.e., in the vicinity of the "normal" level of 4.0 meq/l.

**Table 3 : Fixed effects estimates summarized by the mean(standard error) from the direct h-lik implementation for the 1% , 10% and 100% dataset with random effects at both the IP and HCF levels.**

| Parameter | 1% | 10% | 100% |
|---|---|---|---|
| **Fixed Effects** | | | |
| Intercept | -6.22(0.99) | -5.22(0.31) | -3.21(0.16) |
| Potassium (natural spline coefficient 1) | 6.81(2.46) | 8.65(0.74) | 11.98(0.32) |
| Potassium (natural spline coefficient 2) | -9.49(40.24) | -39.56(12.39) | -77.79(7.97) |
| Potassium (natural spline coefficient 3) | -18.5(79.78) | -74.64(24.57) | -146.92(15.84) |
| eGFR (natural spline coefficient 1) | -0.43(0.25) | -0.66(0.08) | -0.59(0.02) |
| eGFR (natural spline coefficient 2) | -1.84(0.51) | -1.41(0.15) | -1.44(0.05) |
| eGFR (natural spline coefficient 3) | -1.09(0.51) | -0.32(0.15) | -0.26(0.05) |
| Age (natural spline coefficient 1) | 0.31(0.33) | 0.6(0.12) | 0.78(0.04) |
| Age (natural spline coefficient 2) | -0.23(0.91) | 1.43(0.32) | 2.08(0.11) |
| Age (natural spline coefficient 3) | 0.91(0.2) | 1.1(0.07) | 1.22(0.02) |
| Male Gender | 0.21(0.09) | 0.17(0.03) | 0.17(0.01) |
| White Race | 0.16(0.15) | 0.06(0.05) | 0.1(0.02) |
| Hispanic Race | -0.46(0.72) | 0.14(0.15) | 0.16(0.05) |
| Native American | 0.03(0.73) | -0.08(0.2) | 0.16(0.06) |
| Other Race | -0.58(0.32) | -0.12(0.09) | 0.08(0.03) |
| Unknown Race | -0.24(0.47) | -0.09(0.15) | 0.02(0.05) |
| Inpatient status | 2.39(0.19) | 2.13(0.06) | 2.21(0.02) |
| Charlson Comorbidity Score | 0.10(0.01) | 0.10(0.00) | 0.10(0.00) |

**Analysis of performance of *h-lik* computations in large datasets with bivariate and partially crossed random effects**

Total wall time (in log10 space) for triplicate executions of the parallel and serial versions of the proposed implementations of *h-lik* are shown in

**Figure 9** and clearly illustrate the exponential increase in execution time with the size of the dataset irrespective of the machine used. ANOVA identified the size of the dataset (F statistic 1950.474) and the parallel implementation (F statistic 78.030) as more influential predictors than the machine executing the code (F statistic 2.694). A linear regression analysis of the data showed that execution on the Xeon was 16% faster than the i7-5960x (95% confidence interval -38.4% to +2.8% p=0.114). The serial code was 11.5% faster (95% confidence interval -25.8% to +5.6%, p=0.129) slower to execute than the parallel code. In **Figure 10** we illustrate the time spent in various activities when running the *h-lik* programs. Optimization of the $p_{u,\beta}(h)$ profile likelihood took the longest, followed closely by optimization of $p_u(h_{\hat{\gamma}})$. These time-consuming tasks only slightly benefited from parallelization. Tape generation was notably slower when done in parallel. The divergent effects of parallelization on tape generation and optimization, underline the small total effect of parallelization on execution speed.

## Discussion

In this report we demonstrate the initial feasibility of fitting GLMMs to mine large clinical EHR databases by combining the LA with AD into a direct implementation of the *h-lik* approach to GLMMs. Analysis of very large datasets is an area in which many existing approaches to numerical integration for GLMM, e.g., those based on higher order AGH or MCMC methods, are impractical because they execute slowly for big datasets with large number of random effects. In contrast, the proposed technical advance can reliably and reasonably fast fit such datasets. Our implementation differs from the original iGLM algorithm for fitting the hierarchical likelihood. The direct implementation of adjusted profile likelihoods avoids programming errors in the hand coding of matrix operations, as well as numerical errors when calculating derivatives by finite differences speeding up calculations. There are several implications for future work and practice as noted below.

**Random effects modeling in very big datasets.**

GLMMs are one of the most versatile modeling techniques in statistics. In recent years, it has been noted that numerous models used for the non-parametric, flexible modeling of data can be viewed as special cases of the GLMMs. Such models include penalized spline approaches in statistics, but also "workhorses" of Machine Learning such as kernel based methods [62]. However, the applications of GLMMs to big data to date has been limited by the computational bottlenecks inherent in evaluating the high dimensional integrals that arise in their mathematical formulation. A previous biomedically oriented paper reported difficulties in fitting random effects models with 890,934 random effects using AGH in glmer in 2012. The authors resorted to a "divide and conquer" technique to fit separate models to chunks of this dataset followed by meta-regression to synthesize evidence from the smaller datasets. However, because of their approximative nature, any "divide and conquer" technique will be accompanied by a non-negligible amount of bias. Furthermore, certain approximation approaches may exhibit a dramatic loss in statistical efficiency, discarding information and essentially converting a "big", information-rich dataset to a much smaller one [7]. Having the ability to fit a large dataset with the same rigorous techniques that work in small datasets would thus represent a major advantage, because of the preservation of statistical efficiency. In our analyses we effortlessly fit a problem that was 4 times as large as the largest one reported in the biomedical literature by fitting the GLMM using the *h-lik*. Although GLMMs are frequently encountered in the analyses of biomedical data, their scope of application is rather broad. For example, an implementation of GLMMs by the LinkedIn software engineers [9] showed that the random effect modeling performed among the top 5 to 10 methods in numerous nonmedical datasets used in the KDD Cup competitions. Taken together these results suggest that GLMMs deserve a closer examination by other fields that deal with massive datasets using various ad hoc Machine Learning or Artificial Intelligence approaches.

**H-lik offers a theoretically rigorous, computational friendly method for estimation of GLMMs.**

The *h-lik* was originally introduced as a unified framework for several common models e.g. Poisson-gamma, binomial-beta with the intention to resolve apparent differences between subject specific and population averaged models[10]. *h-lik* was latter extended to models with structured and unknown dispersion parameters (e.g. the (inverse) Gaussian GLMM) [11] and even the so called "double HGLM" in which random effects can be specified for both mean and variance parameters. The reliance of *h-lik* on the LA would appear to be an area of concern, especially in datasets similar to our own, with few repeated measurements per individual and a binary outcome[35, 36, 63]. Our simulations show that these concerns may not be as limiting as one may think, especially when modeling takes place in the log-relative risk, rather than the logit scale. In both simulated and CRWD datasets, the LA based *h-lik* inferences for fixed and random effects were numerically close to those obtained via the alternative AGH and MCMC approaches. In the case of varying estimates between *h-lik* and AGH, we found the former to be closer to the gold standard of MCMC integration. This observation suggests that higher order LA introduced in extensions of the basic *h-lik* theory [11, 12, 32, 35, 36] may not be required for the size of datasets considered in this paper. At first, this conclusion appears to contradict previous literature arguing that higher order LA are required to avoid biases in the estimation of variance components[32, 36, 64–68]. However, this apparent contradiction disappears if one considers the asymptotic order behavior of the LA to integrals over *very high* dimensional spaces[69]. The error (and thus potential inferential bias) introduced by foregoing corrections is of the order of $o(k^{-(1+d)/2})$, where $d$ is the dimensionality of the space and $k$ a scaling factor of the Hessian at the optimum.

Therefore, for analyses of datasets with large number of random effects (large $d$), in which the underlying model is reasonably identified by the data (large $k$), the approximation error of the LA will rapidly decline to zero because of the "little o" asymptotics. Consequently, while the curse of dimensionality renders higher order AGH approximations impractical, it simultaneously makes the ordinary LA highly accurate in the same settings, at least for non-ultra-sparse datasets. Even in that case, the proposed implementation of the *h-lik* corresponds to method HL(1,1) in the theory of the hierarchical likelihood, which appears to not be as susceptible to bias as lower order approximations [32, 36, 52, 70] for sparse binary data.

The *h-lik* approach may be viewed as Empirical Bayes/modal approximations to a non-informative Bayesian analysis of the same data, allowing the analyst to invoke a Bayesian interpretation of the results obtained through the *h-lik* if non-informative priors are appropriate for the problem at hand. In the opinion of the authors, this is a rather significant development since bona fide Bayesian analyses of large datasets via MCMC can be orders of magnitude slower than their likelihood counterparts. Future studies should explore the use (and abuse) of *h-lik* as approximations to full Bayesian analyses in other datasets and define the range of problems and random effects structures for which *h-lik* will give numerical answers that do not differ much from those given by MCMC. A Bayesian "repurposing" of the *h-lik* could thus allow a wider adoption of Bayesian methods. A particularly interesting extension would also explore non-Gaussian random effect distributions, which are trivial to estimate both by the iGLM algorithm and the implementation considered herein.

Scalable alternatives to the proposed *h-lik* implementation, include the R package glmmTMB and the original iGLM algorithm for fitting the hierarchical likelihood approach. glmmTMB was recently introduced [53] for GLMMs and uses an interface similar to R's glmer. glmmTMB offers a "REML" option, yet the precise definition of the REML implemented is not explicitly stated in the package documentation. Examination of the R source code of the glmmTMB package suggests that REML estimation amounts to optimization of the $p_{u,\beta}(h)$ function, i.e., the first stage in the hierarchical likelihood approach. We have verified this to be the case by contrasting the glmmTMB fixed effects estimates to those returned by the first optimization in the *h-lik* approach. Limited comparisons in subsets of CRWD and simulations showed that while the variance components returned by glmmTMB and our approach were identical, estimates of the fixed and random effects could differ, particularly when the number of random effects in the dataset were small. Future work should thus concentrate on investigating alternative ways to optimize the adjusted profile likelihood functions of *h-lik* theory that can offer higher speed advantages to our direct implementation.

The original iGLM approach offers an alternative to the direct implementation of the *h-lik* approach with comparable performance in small datasets. This alternative is also a candidate for a high-performance implementation, since the underlaying operations (iterative solutions of weighted linear regression and numerical optimization) are available in many high-performance scientific computing platforms. Future work may even consider implementing either of the two implementations of the *h-lik* explored in this work within mainstream Machine Learning platforms such as TensorFlow[71]. R research in alternative implementations of the *h-lik* thus has the potential to make the rigor of GLMMs available to a wider audience of data scientists in addition to statisticians.

**What did we learn about the potassium level and EHR analyses?**

This work was motivated by a clinical question that the authors posed internally a few years ago. We believe that the analysis presented herein answered the question by re-confirming the "U" shaped curve between the K$^+$ level and mortality previously suggested [14], but dramatically reducing the uncertainty about this relationship. Based on our data, potassium levels close to the normal level of 4meq/l are not associated with excess mortality. Hence, our bracketing of a globally valid reference range for potassium levels could be used to inform the design of *clinical trials* to guide the proper use of potassium supplements and potassium lowering drugs. The individual and facility level variation detected here is of relevance to future investigations that use EHRs; unless such variation can be captured and modelled through covariates available in the EHRs, inferences from such data will be contaminated by significant bias.

## Conclusions

Combining the Laplace Approximation with Automatic Differentiation results in a direct implementation of the *h-lik* that can efficiently fit very large EHR datasets, with accuracy that is equivalent if not better than the state-of-the-art AGH methods and the original iGLM algorithm for GLMMs. For the problems we examined, the h-lik gave results that are indistinguishable from the gold standard of MCMC integration, but the results were obtained much faster.

## List of abbreviations:

**AD:** Algorithmic Differentiation

**AGH(x):** Adaptive Gaussian Hermite Quadrature (x: numeral indicating the order of the method)

**ANOVA:** Analysis of Variance

**BLAS:** Basic Linear Algebra Subprograms

**BLUP**: Best Linear Unbiased Predictor

**CRWD:** Cerner Real World Data

**eGFR:** Estimated Glomerular Filtration Rate

**EHR:** Electronic Health Record

**EQL:** Extended Quasi Likelihood

**GLM:** Generalized Linear Model

**GLMM:** Generalized Linear Mixed Model

**HCP:** Healthcare Practitioners

**HCF:** Healthcare Facilities

**h-lik:** Hierarchical Likelihood

**K$^+$:** Serum potassium level

**iGLM:** Interconnected Generalized Linear Models

**IWLS:** Iterative Weighted Linear Squares

**IP:** Individual Patients

**IQR:** Inter-Quartile Range

**LA:** Laplace Approximation

**MCMC:** Markov Chain Monte Carlo

**MLE:** Maximum Likelihood Estimation

**REML:** Restricted Maximum Likelihood

## Declarations:

**Ethics approval and consent to participate:** Analyses with CRWD are considered "Not Human Research" by the Human Research Protection Office (ethics committee and Institutional Review Board) of the University of New Mexico Health Sciences Center

**Consent for publication:** Not applicable

**Availability of data and materials:** The data that support the findings of this study are available from Cerner Corporation, but restrictions apply to the availability of these data, which were used under license for the current study, and so are not publicly available. The Cerner Health Facts Database (now referred to as the Cerner Real World Data, CRWD) is available to researchers at contributing hospitals (and their research affiliates) upon request made directly to Cerner Corporation. Software code for the implementation of the LA for a Poisson regression with one and two random effects is available from the repository: https://bitbucket.org/chrisarg/laplaceapproximationandhyperkalemia/

**Competing interests:** CA has received consulting fees from Bayer, Baxter Healthcare, Health Services Advisory Group and research support from Dialysis Clinic, Inc. CGB, VSP MLU, MER, VS, SKS, SA have received support from Dialysis Clinic, Inc. JC has nothing to disclose.

**Funding:** This research was funded in part by Dialysis Clinic Inc under grant RF#C-4011

**Author's contributions:** CB extracted and preprocessed the data from the CRWD database used in this analysis, VSP, JC and CA designed the comparative evaluations used in the manuscript and programmed the analysis in R. MLU, MER, VS, SKS, SA and CA interpreted the clinical analyses of hyperkalemia and the risk of death.

**Acknowledgements:** Not applicable

# References:

1. Bates DW, Saria S, Ohno-Machado L, Shah A, Escobar G. Big Data In Health Care: Using Analytics To Identify And Manage High-Risk And High-Cost Patients. Health Affairs. 2014;33:1123–31.

2. Goldstein BA, Navar AM, Pencina MJ, Ioannidis JPA. Opportunities and challenges in developing risk prediction models with electronic health records data: a systematic review. J Am Med Inform Assoc. 2017;24:198–208.

3. Krumholz HM. Big Data And New Knowledge In Medicine: The Thinking, Training, And Tools Needed For A Learning Health System. Health Affairs. 2014;33:1163–70.

4. Silverio A, Cavallo P, De Rosa R, Galasso G. Big Health Data and Cardiovascular Diseases: A Challenge for Research, an Opportunity for Clinical Care. Front Med (Lausanne). 2019;6. doi:10.3389/fmed.2019.00036.

5. Rajkomar A, Oren E, Chen K, Dai AM, Hajaj N, Hardt M, et al. Scalable and accurate deep learning with electronic health records. npj Digital Med. 2018;1:1–10.

6. Gebregziabher M, Egede L, Gilbert GE, Hunt K, Nietert PJ, Mauldin P. Fitting parametric random effects models in very large data sets with application to VHA national data. BMC Medical Research Methodology. 2012;12:163.

7. Perry PO. Fast moment-based estimation for hierarchical models. Journal of the Royal Statistical Society: Series B (Statistical Methodology). 2017;79:267–91.

8. Lee JYL, Brown JJ, Ryan LM. Sufficiency Revisited: Rethinking Statistical Algorithms in the Big Data Era. The American Statistician. 2017;71:202–8.

9. Zhang X, Zhou Y, Ma Y, Chen B-C, Zhang L, Agarwal D. GLMix: Generalized Linear Mixed Models For Large-Scale Response Prediction. In: Proceedings of the 22Nd ACM SIGKDD International Conference on Knowledge Discovery and Data Mining. New York, NY, USA: ACM; 2016. p. 363–72. doi:10.1145/2939672.2939684.

10. Lee Y, Nelder JA. Hierarchical Generalized Linear Models. Journal of the Royal Statistical Society Series B (Methodological). 1996;58:619–78.

11. Lee Y, Nelder JA. Hierarchical Generalised Linear Models: A Synthesis of Generalised Linear Models, Random-Effect Models and Structured Dispersions. Biometrika. 2001;88:987–1006.

12. Lee Y, Nelder JA, Pawitan Y. Generalized Linear Models with Random Effects: Unified Analysis via H-likelihood, Second Edition. 2 edition. Boca Raton, Florida: Chapman and Hall/CRC; 2017.

13. Nilsson E, Gasparini A, Ärnlöv J, Xu H, Henriksson KM, Coresh J, et al. Incidence and determinants of hyperkalemia and hypokalemia in a large healthcare system. International Journal of Cardiology. 2017;245:277–84.

14. Luo J, Brunelli SM, Jensen DE, Yang A. Association between Serum Potassium and Outcomes in Patients with Reduced Kidney Function. CJASN. 2016;11:90–100.

15. Chapman Neil, Dobson Joanna, Wilson Sarah, Dahlöf Björn, Sever Peter S., Wedel Hans, et al. Effect of Spironolactone on Blood Pressure in Subjects With Resistant Hypertension. Hypertension. 2007;49:839–45.

16. Zannad F, McMurray JJV, Krum H, van Veldhuisen DJ, Swedberg K, Shi H, et al. Eplerenone in Patients with Systolic Heart Failure and Mild Symptoms. http://dx.doi.org/10.1056/NEJMoa1009492. 2011. doi:10.1056/NEJMoa1009492.

17. Pitt B, Zannad F, Remme WJ, Cody R, Castaigne A, Perez A, et al. The Effect of Spironolactone on Morbidity and Mortality in Patients with Severe Heart Failure. http://dx.doi.org/10.1056/NEJM199909023411001. 2008. doi:10.1056/NEJM199909023411001.

18. Linde C, Qin L, Bakhai A, Furuland H, Evans M, Ayoubkhani D, et al. Serum potassium and clinical outcomes in heart failure patients: results of risk calculations in 21 334 patients in the UK. ESC Heart Fail. 2019;6:280–90.

19. Bakris GL, Agarwal R, Anker SD, Pitt B, Ruilope LM, Rossing P, et al. Effect of Finerenone on Chronic Kidney Disease Outcomes in Type 2 Diabetes. New England Journal of Medicine. 2020;0:null.

20. Bolignano D, Palmer SC, Navaneethan SD, Strippoli GFM. Aldosterone antagonists for preventing the progression of chronic kidney disease. Cochrane Database Syst Rev. 2014;:CD007004.

21. Trevisan M, de Deco P, Xu H, Evans M, Lindholm B, Bellocco R, et al. Incidence, predictors and clinical management of hyperkalaemia in new users of mineralocorticoid receptor antagonists. Eur J Heart Fail. 2018;20:1217–26.

22. Bologa C, Pankratz VS, Unruh ML, Roumelioti ME, Shah V, Shaffi SK, et al. Generalized Mixed Modeling in Massive Electronic Health Record Databases: what is a healthy serum potassium? arXiv:191008179 [stat]. 2019. http://arxiv.org/abs/1910.08179. Accessed 16 Nov 2020.

23. Argyropoulos C, Bologa C George, Pankratz VS, Unruh ML, Roumelioti ME, Shah V, et al. Association of Potassium Level and Mortality in Massive Health Record Databases. In: Journal of The American Society of Nephrology. San Diego; 2018. p. 499–500.

24. Argyropoulos C, Unruh ML. Analysis of Time to Event Outcomes in Randomized Controlled Trials by Generalized Additive Models. PLoS ONE. 2015;10:e0123784.

25. Liu Q, Pierce DA. A Note on Gauss-Hermite Quadrature. Biometrika. 1994;81:624–9.

26. Pinheiro JC, Bates DM. Approximations to the Log-Likelihood Function in the Nonlinear Mixed-Effects Model. Journal of Computational and Graphical Statistics. 1995;4:12–35.

27. Pinheiro JC, Bates DM. Mixed Effects Models in S and S-Plus. Springer; 2000.

28. Pinheiro JC, Chao EC. Efficient Laplacian and Adaptive Gaussian Quadrature Algorithms for Multilevel Generalized Linear Mixed Models. Journal of Computational and Graphical Statistics. 2006;15:58–81.

29. Liu Q. Laplace approximations to likelihood functions for generalized linear mixed models. 1994. https://ir.library.oregonstate.edu/concern/graduate_thesis_or_dissertations/b8515r59n?locale=en.

30. Wolfinger R. Laplace's Approximation for Nonlinear Mixed Models. Biometrika. 1993;80:791–5.

31. Skinner L. Note on the Asymptotic Behavior of Multidimensional Laplace Integrals. SIAM J Math Anal. 1980;11:911–7.

32. Collins D. The performance of estimation methods for generalized linear mixed models. University of Wollongong; 2008. https://ro.uow.edu.au/theses/1737.

33. Capanu M, Gönen M, Begg CB. An assessment of estimation methods for generalized linear mixed models with binary outcomes. Statist Med. 2013;32:4550–66.

34. McGilchrist CA, Yau KKW. The derivation of blup, ML, REML estimation methods for generalised linear mixed models. Communications in Statistics - Theory and Methods. 1995;24:2963–80.

35. Noh M, Lee Y. REML estimation for binary data in GLMMs. Journal of Multivariate Analysis. 2007;98:896–915.

36. Lee W, Lee Y. Modifications of REML algorithm for HGLMs. Stat Comput. 2012;22:959–66.

37. Cox DR, Reid N. Parameter Orthogonality and Approximate Conditional Inference. Journal of the Royal Statistical Society: Series B (Methodological). 1987;49:1–18.

38. Lee Y, Nelder JA. Double hierarchical generalized linear models (with discussion). Journal of the Royal Statistical Society: Series C (Applied Statistics). 2006;55:139–85.

39. Nelder JA, Pregibon D. An Extended Quasi-Likelihood Function. Biometrika. 1987;74:221–32.

40. Rönnegård L, Shen X, Alam M. hglm: A Package for Fitting Hierarchical Generalized Linear Models. The R Journal. 2010;2:20–8.

41. Molas M, Lesaffre E. Hierarchical Generalized Linear Models: The R Package HGLMMM. Journal of Statistical Software. 2011;39:1–20.

42. Beck A, Tetruashvili L. On the Convergence of Block Coordinate Descent Type Methods. SIAM J Optim. 2013;23:2037–60.

43. Lange K, Chi EC, Zhou H. A Brief Survey of Modern Optimization for Statisticians: Modern Optimization for Statisticians. International Statistical Review. 2014;82:46–70.

44. Griewank Andreas, Walther Andrea. Evaluating Derivatives. Society for Industrial and Applied Mathematics; 2008. doi:10.1137/1.9780898717761.

45. Bartholomew-Biggs M, Brown S, Christianson B, Dixon L. Automatic differentiation of algorithms. Journal of Computational and Applied Mathematics. 2000;124:171–90.

46. Baydin AG, Pearlmutter BA, Radul AA, Siskind JM. Automatic Differentiation in Machine Learning: a Survey. Journal of Machine Learning Research. 2018;18:1–43.

47. Gebremedhin AH, Manne F, Pothen A. What Color Is Your Jacobian? Graph Coloring for Computing Derivatives. SIAM Review. 2005;47:629–705.

48. Coleman TF, Moré JJ. Estimation of sparse hessian matrices and graph coloring problems. Mathematical Programming. 1984;28:243–70.

49. Skaug HJ. Automatic Differentiation to Facilitate Maximum Likelihood Estimation in Nonlinear Random Effects Models. Journal of Computational and Graphical Statistics. 2002;11:458–70.

50. Skaug HJ, Fournier DA. Automatic approximation of the marginal likelihood in non-Gaussian hierarchical models. Computational Statistics & Data Analysis. 2006;51:699–709.

51. Kristensen K, Nielsen A, Berg CW, Skaug H, Bell BM. TMB: Automatic Differentiation and Laplace Approximation. Journal of Statistical Software. 2016;70. doi:10.18637/jss.v070.i05.

52. Yun S, Lee Y. Comparison of hierarchical and marginal likelihood estimators for binary outcomes. Computational Statistics & Data Analysis. 2004;45:639–50.

53. Brooks ME, Kristensen K, Benthem KJ van, Magnusson A, Berg CW, Nielsen A, et al. glmmTMB Balances Speed and Flexibility Among Packages for Zero-inflated Generalized Linear Mixed Modeling. The R Journal. 2017;9:378–400.

54. Blackford LS, Petitet A, Pozo R, Remington K, Whaley RC, Demmel J, et al. An updated set of basic linear algebra subprograms (BLAS). ACM Transactions on Mathematical Software. 2002;28:135–51.

55. Meng X-L. Decoding the H-likelihood. Statist Sci. 2009;24:280–93.

56. Bender A, Groll A, Scheipl F. A generalized additive model approach to time-to-event analysis. Statistical Modelling. 2018;:1471082X17748083.

57. Burton A, Altman DG, Royston P, Holder RL. The design of simulation studies in medical statistics. Stat Med. 2006;25:4279–92.

58. Bates D, Mächler M, Bolker B, Walker S. Fitting Linear Mixed-Effects Models Using lme4. Journal of Statistical Software. 2015;67:1–48.

59. Stan Development Team. Stan Modeling Language Users Guide and Reference Manual, Version 2.25. 2019. http://mc-stan.org/.

60. Stan Development Team. RStan: the R interface to Stan. 2020. http://mc-stan.org/.

61. Hoffman MD, Gelman A. The No-U-Turn Sampler: Adaptively Setting Path Lengths in Hamiltonian Monte Carlo. Journal of Machine Learning Research. 2014;15:1593–623.

62. Ruppert D, Wand MP, Carroll RJ. Semiparametric regression during 2003–2007. Electron J Stat. 2009;3:1193–256.

63. Vonesh EF. A Note on the Use of Laplace's Approximation for Nonlinear Mixed-Effects Models. Biometrika. 1996;83:447–52.

64. Breslow NE, Lin X. Bias Correction in Generalised Linear Mixed Models with a Single Component of Dispersion. Biometrika. 1995;82:81–91.

65. Lin X, Breslow NE. Bias Correction in Generalized Linear Mixed Models With Multiple Components of Dispersion. Journal of the American Statistical Association. 1996;91:1007–16.

66. Shun Z. Another Look at the Salamander Mating Data: A Modified Laplace Approximation Approach. Journal of the American Statistical Association. 1997;92:341–9.

67. Shun Z, McCullagh P. Laplace Approximation of High Dimensional Integrals. Journal of the Royal Statistical Society Series B (Methodological). 1995;57:749–60.

68. Noh M, Lee Y. REML estimation for binary data in GLMMs. Journal of Multivariate Analysis. 2007;98:896–915.

69. Kirwin WD. Higher asymptotics of Laplace's approximation. Asymptotic Analysis. 2010;70:231–48.

70. Lee W, Lim J, Lee Y, del Castillo J. The hierarchical-likelihood approach to autoregressive stochastic volatility models. Computational Statistics & Data Analysis. 2011;55:248–60.

71. Martín Abadi, Ashish Agarwal, Paul Barham, Eugene Brevdo, Zhifeng Chen, Craig Citro, et al. TensorFlow: Large-Scale Machine Learning on Heterogeneous Systems. 2015. https://www.tensorflow.org/.

# Figure Legends

**Figure 1:** Performance of GLMM fitting methods in the simulated Poisson datasets: fixed effects (A), variance components (B) and random effects (C). For set of parameters we show the standardized Bias (stdBias) and the Mean Square Error (MSE). Abbreviations: AGH0(1) : Adaptive Gaussian Quadrature of order 0 or 1, glmmTMB: estimates returned by the relevant package in R, h-lik: the proposed method in text, iGLM: interconnected Generalized Linear Model, MCMC : Markov Chain Monte Carlo (MCMC) posterior mean , MCMCmode: mode of posterior marginals from MCMC simulations.  N of IP: Number of Individual Patients in each simulated dataset.

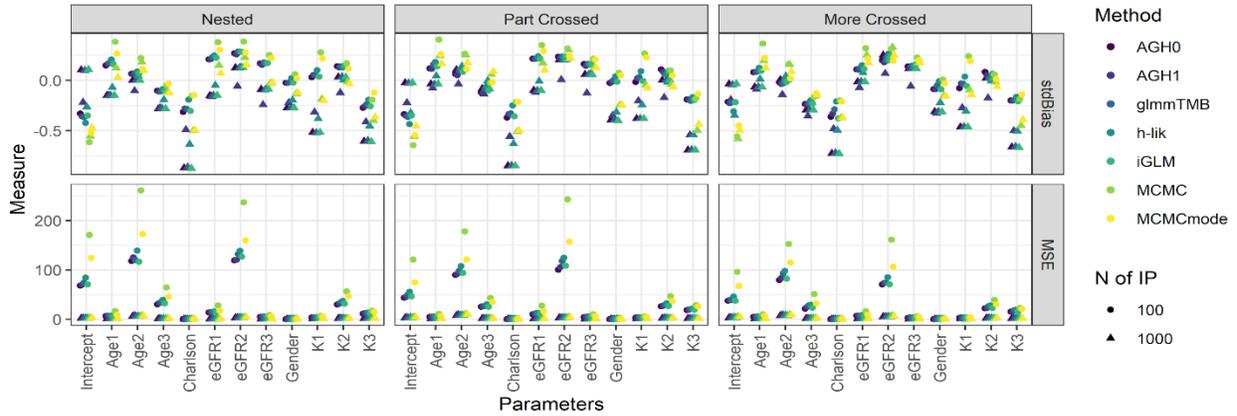

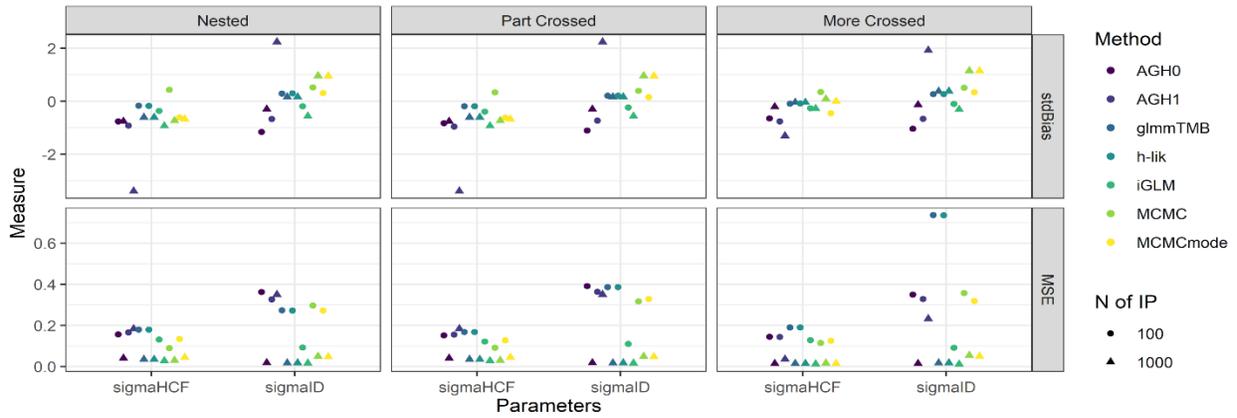

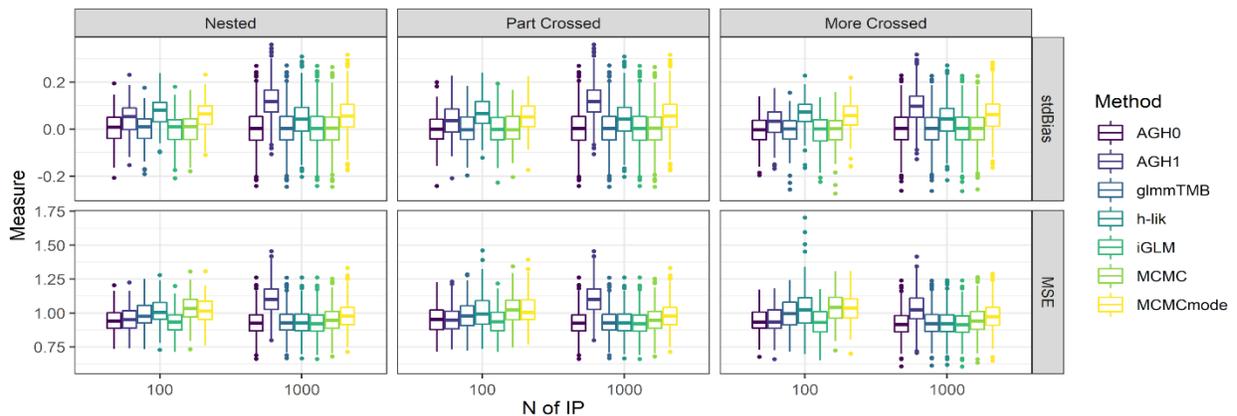

**Figure 2:** Performance of GLMM fitting methods in the less variable Binomial datasets: fixed effects (A), variance components (B) and random effects (C). For set of parameters we show the standardized Bias (stdBias) and the Mean Square Error (MSE). Abbreviations: AGH0(1) : Adaptive Gaussian Quadrature of order 0 or 1, glmmTMB: estimates returned by the relevant package in R, h-lik: the proposed method in text, h-lik-pub: estimates returned by the first optimization for $p_{u,\beta}(h)$ in the proposed implementation, h-lik-3, random effect estimates from the optimization of $h(\beta, u, \hat{\gamma}, \hat{\phi}; Y, u)$ over the fixed and random effects.

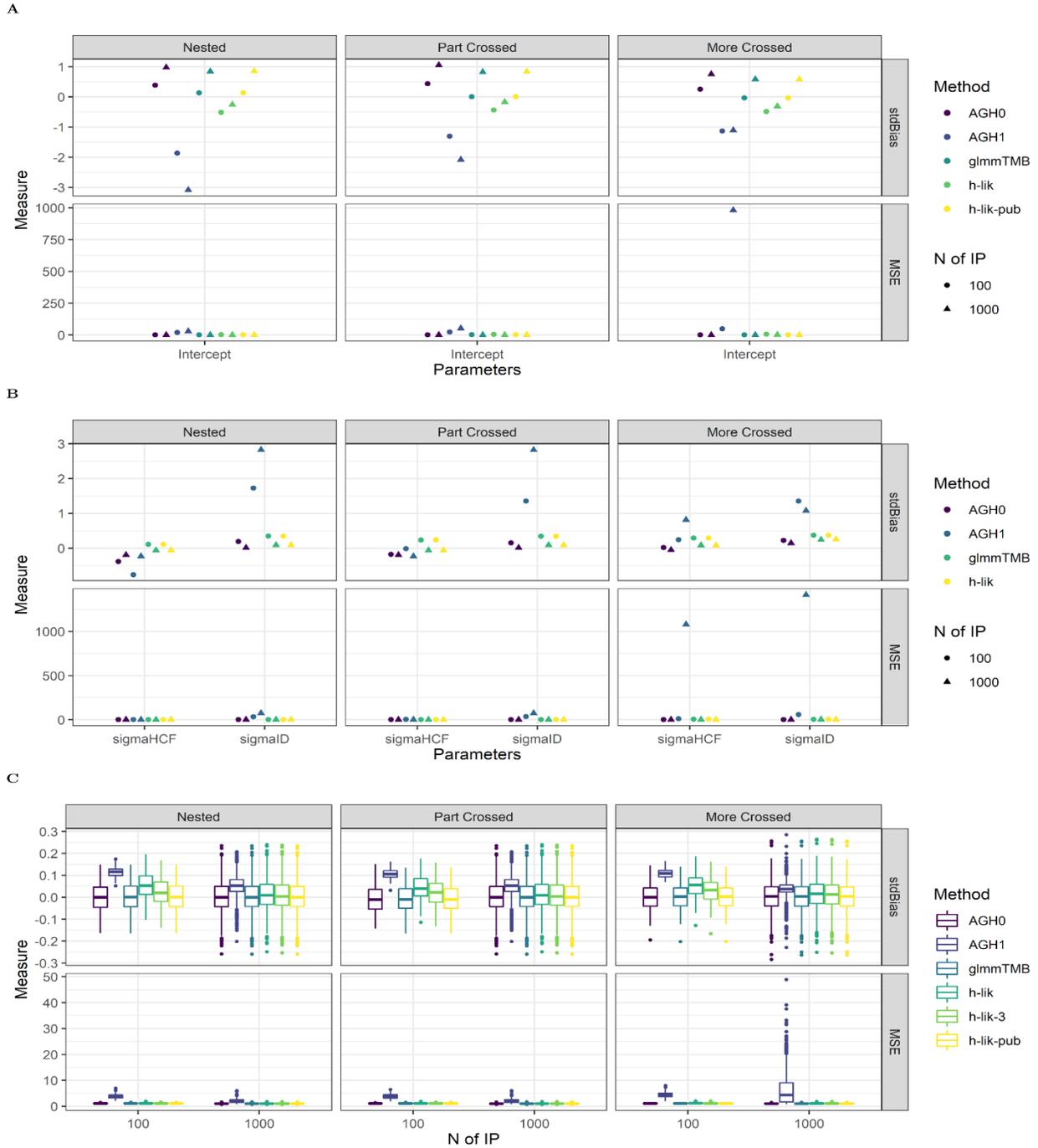

**Figure 3:** Performance of GLMM fitting methods in the More variable Binomial datasets: fixed effects (A), variance components (B) and random effects (C). For set of parameters we show the standardized Bias (stdBias) and the Mean Square Error (MSE). Abbreviations: glmmTMB: estimates returned by the relevant package in R, h-lik: the proposed method in text, h-lik-pub: estimates returned by the first optimization for $p_{u,\beta}(h)$ in the proposed implementation.

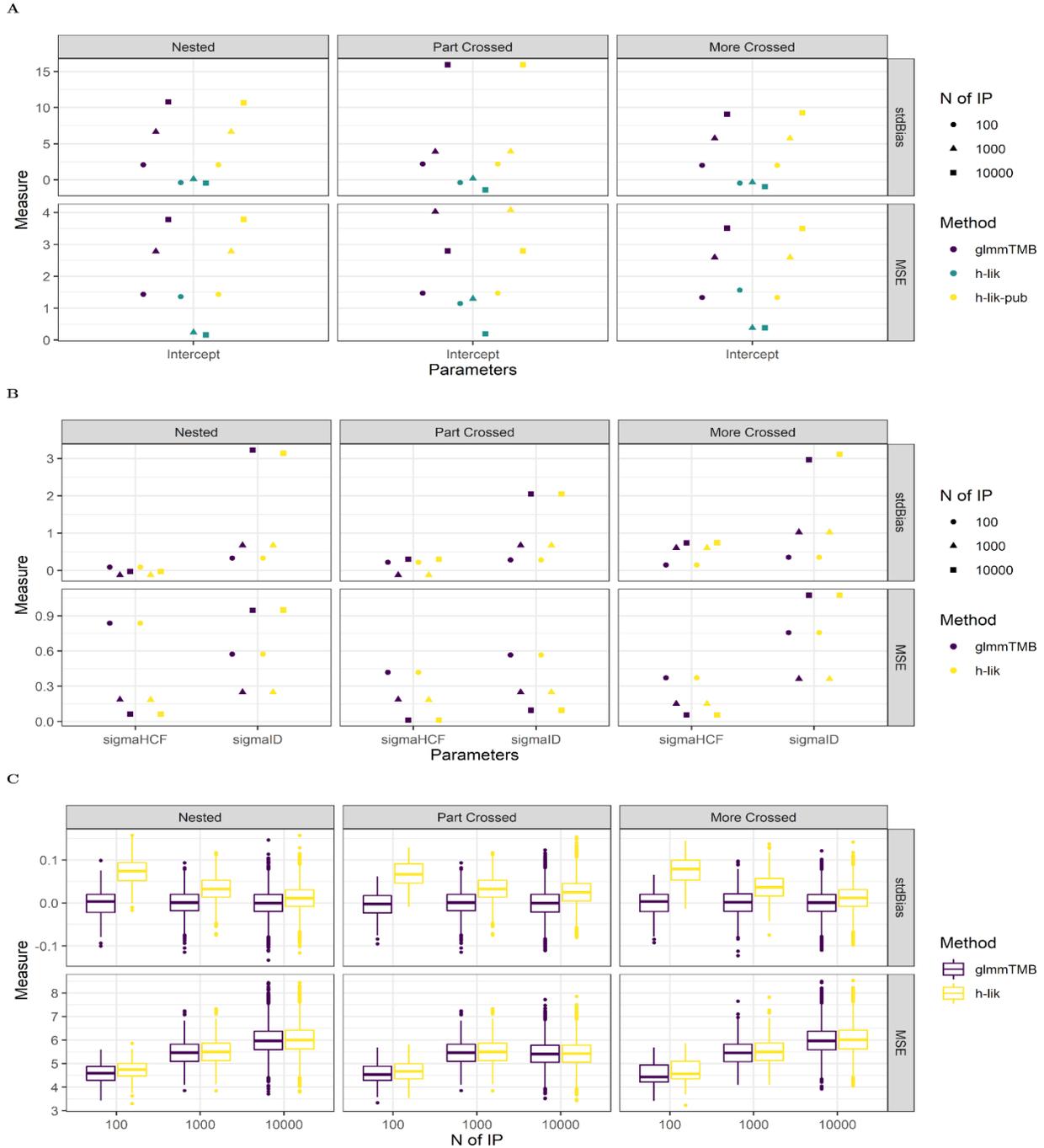

**Figure 4:** Estimates and 95% confidence intervals for the fixed effect components in the CRWD dataset

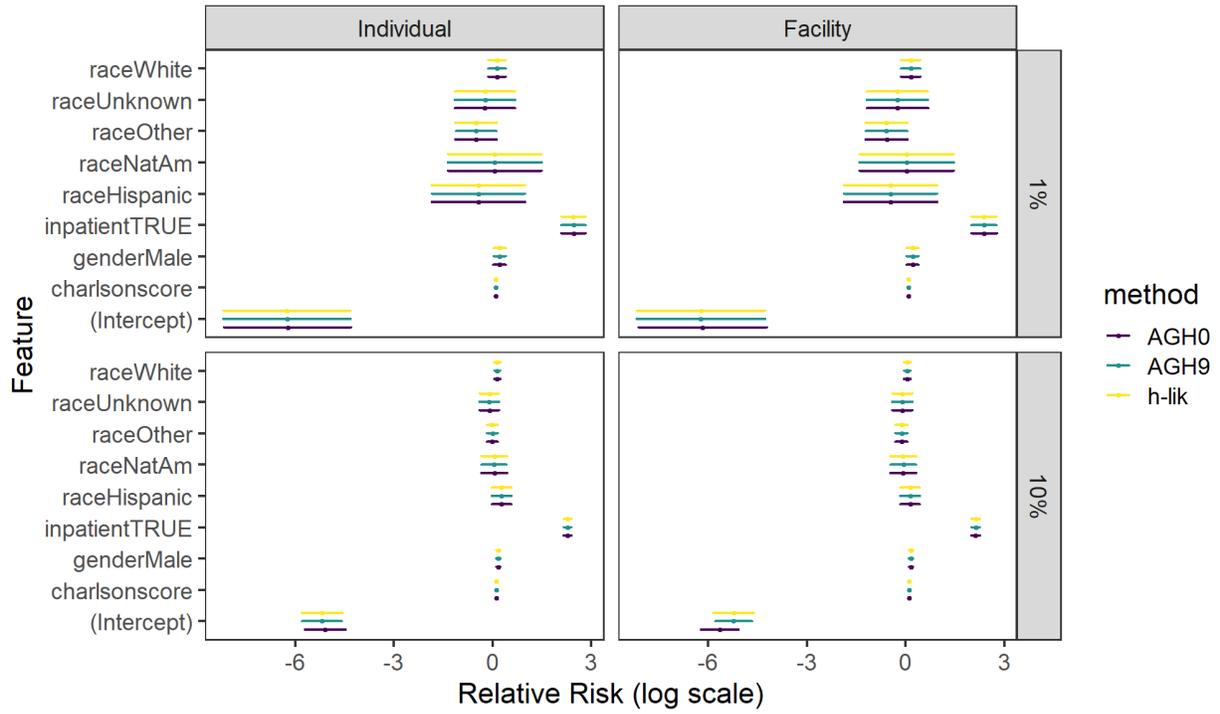

**Figure 5**: Estimated cubic splines relating the relative risk of death to potassium level and their associated 95% confidence (*h-lik*) and credible (MCMC) intervals using the 1% dataset and a single random effect at the healthcare facility level

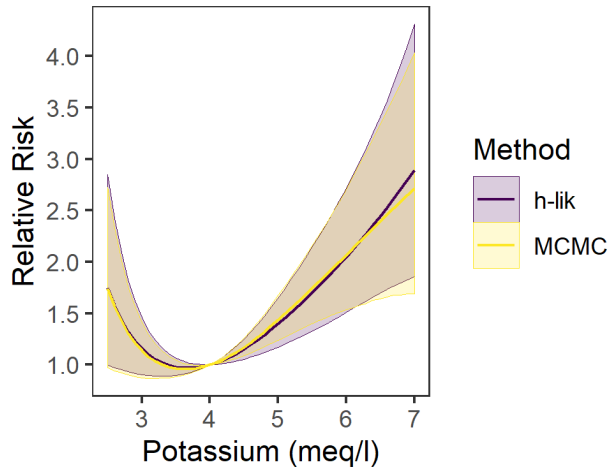

**Figure 6**: Comparison of the predicted random effects (BLUPs) as estimated by the *h*-lik, the various AGH and the MCMC methods

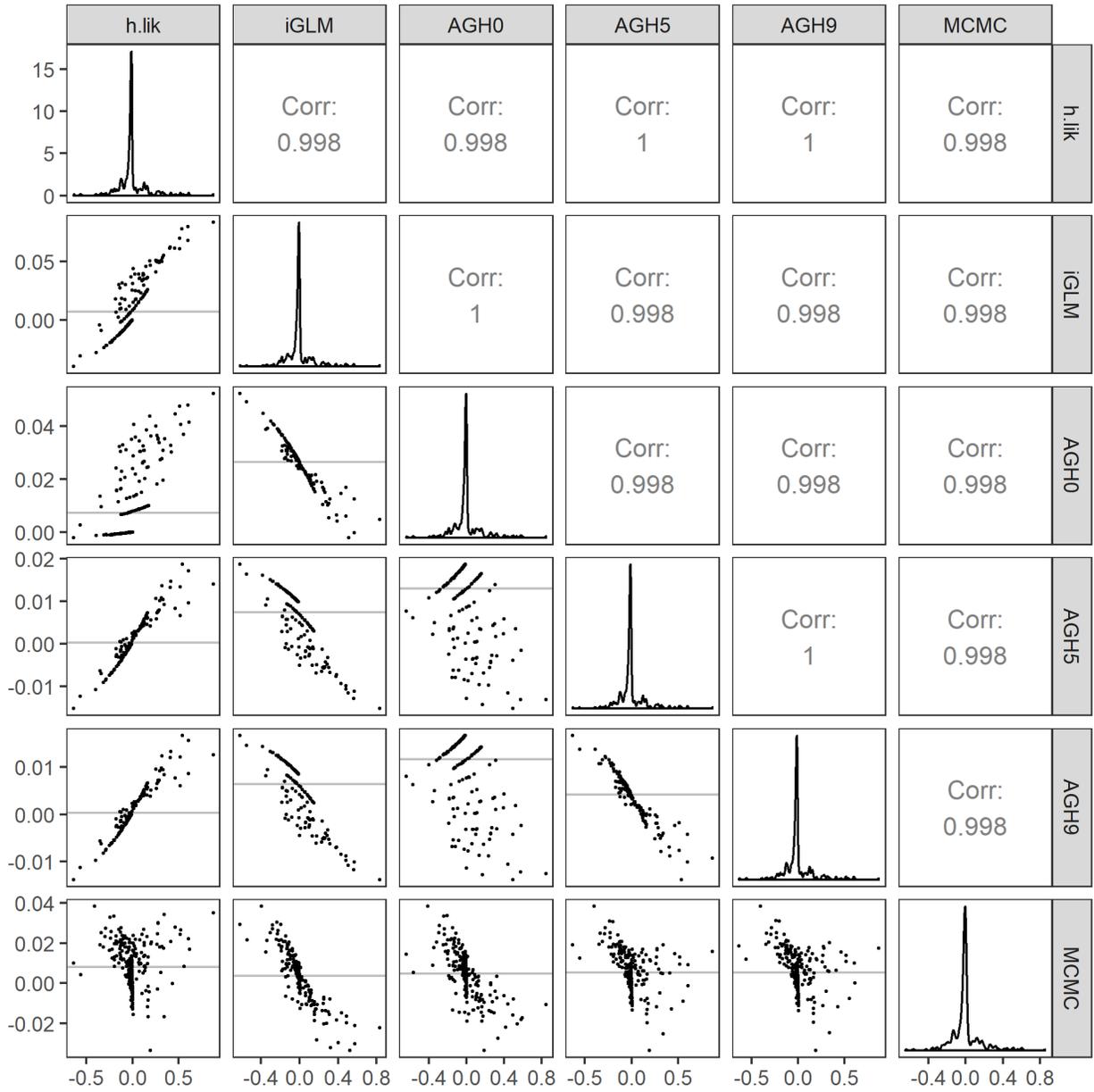

**Figure 7**: GLMM execution timings for different datasets for single random effects models, either at the individual patient or the healthcare facility level

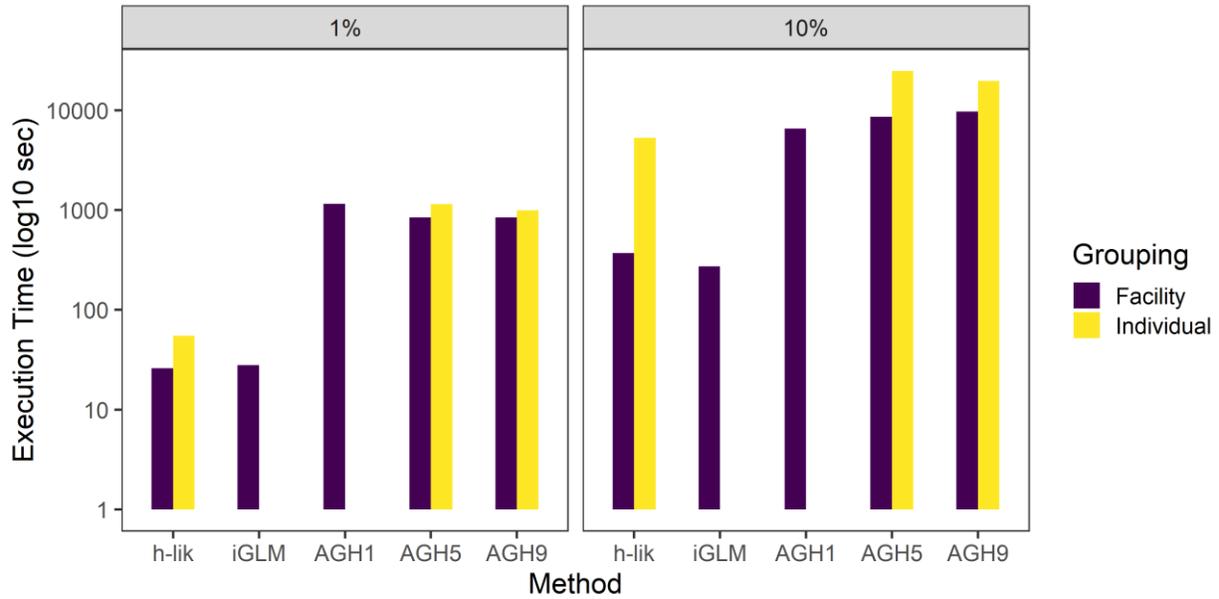

**Figure 8**: Estimated cubic splines relating the relative risk of death to potassium level and their associated 95% confidence by the direct *h-lik* implementation for various dataset sizes. Poisson models included random effects at both the individual and the healthcare facility levels (gray lines demarcate the zone in which RR is between 0.9 and 1.1).

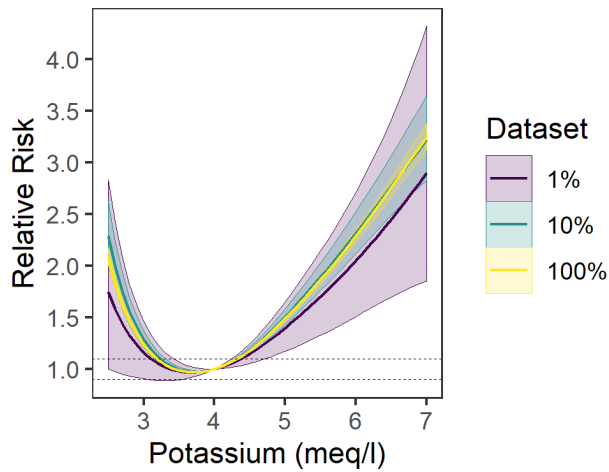

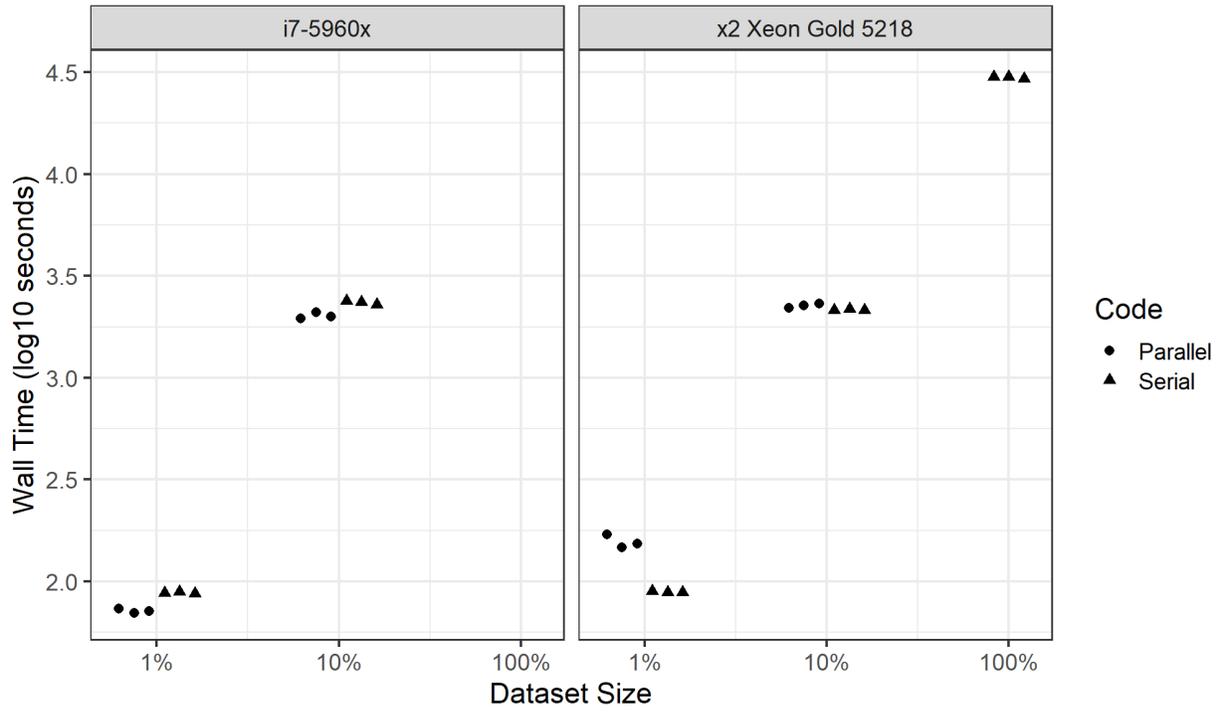

**Figure 9**: Performance (total wall time) of *h-lik* computations using serial and parallel codes.

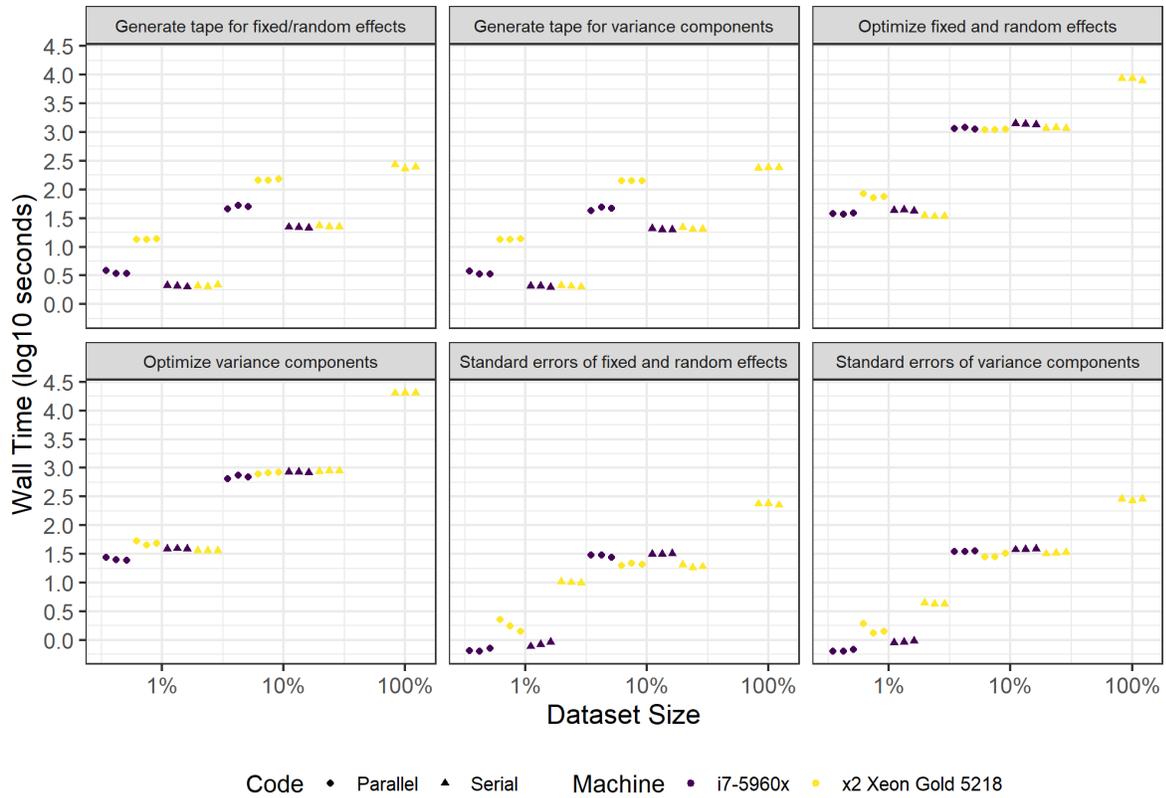

**Figure 10**: Execution times of the steps of *h-lik* calculations for serial and parallel codes.

# Appendix
# Supplementary Methods
## Simulation of Artificial Repeated Measures Electronic Health Record Datasets with Arbitrary Crossed Random Effect Design Matrix Structure

In this section we discuss how we simulated datasets with two random effects that could be either nested or crossed to an arbitrary extent. For Electronic Health Records (EHR), we can conceptualize these random effects to arise from *inter-individual variability*, i.e. random variation between patients that is not adequately captured by fixed covariates and *inter-healthcare-facility* variation, i.e., random variation due to facility policies and procedures. In a different, e.g., educational setting, one could postulate variability due to students and teachers/curricula. Real world datasets of this nature are never balanced since one would never have the same number of observations per cluster; furthermore, they will most often demonstrate an unbalanced crossed structure, i.e., not every lower-level unit will contribute to the same number of upper-level units. By the latter we mean, that the observations from a cluster at the lower level of the hierarchy may provide information for more than one clusters at a higher level of the hierarchy. For example, a patient (lower-level cluster) may visit more than healthcare facilities (higher level cluster), or a student may be taught by more than one educator. However, not every patient or student will interact with all doctors or teachers. To conform to our healthcare focus, we will refer to the lower (higher) level of clusters as "individual patients" (IP) and "healthcare facilities" (HCF) respectively. We devised a hierarchical simulation strategy to generate arbitrary datasets with a partially crossed structure:

1. We fix the number of IP ($N_{IP}$) and HCF ($N_{HCF}$) respectively.
2. For $j = 1, \ldots, N_{IP}$, we simulated observations ("visits") ($N_j^v$) from a truncated negative binomial distribution. The latter was parameterized by a mean parameter ($\mu_v$), a size parameter ($\phi_v$), a lower ($m_v$) and an upper bound ($M_v$), i.e.,

$$N_j^v \sim NB(\mu_v, \phi_v)T(m_v, M_v)$$

   The corresponding non-truncated negative binomial has mean $\mu_v$ and variance $\mu_v + \mu_v^2/\phi_v$. This specification allows us to vary both the expected number of repeated measures per individual ($\mu_v$), and control how unbalanced the dataset ($\phi_v$) will be, while limiting extreme events (number of repeated observations too high, or too low) that may arise during simulations by using the two bounds.

3. For $j = 1, \ldots, N_{IP}$, we simulated the number of distinct HCF ($N_j^f$) the jth patient visited from a truncated Poisson distribution. The latter was parameterized by a rate parameter ($\lambda_f$), a lower ($m_f$) and an upper bound ($M_f$), i.e.,

$$N_j^f \sim Poisson(\lambda_f)T(m_f, M_f)$$

   The rate parameter controls the propensity of each lower unit to be associated with numerous upper-level units, while the use of bounds allows one to further tune this propensity. In the case that $m_f > 0$ and $M_f < 2$, this specification allows the user to simulate nested random effects.

4. For $j = 1, \ldots, N_{IP}$, we simulated the probability that the jth individual visits one of the $N_j^f$ facilities, as well as the number of visits to each of these facilities in hierarchical fashion:
   4.1. Sample positive quantities $\alpha_1, \ldots, a_{N_{HCF}} \sim Lognormal(\mu_\alpha, \sigma_\alpha)$

4.2. Sample uniformly and without replacement $z_1, \ldots, z_{N_j^v}$ from the set $\alpha_1, \ldots, \alpha_{N_{HCF}}$

4.3. Calculate the probability that the jth individual will visit each of the $1, \ldots, N_j^f$ as:

$$p_{i,j} = \frac{z_i}{\sum_{k=1}^{N_j^f} z_k}$$

4.4. Sample the number of visits $n_{1,j}, \ldots, n_{N_j^v,j}$ from the multinomial distribution with total sample size $N_j^f$ and vector of probabilities $p_{i,j}, i = 1, \ldots, N_j^f$:

$$n_{1,j}, \ldots, n_{N_j^v,j} \sim Multinomial\left(N_j^f; p_{1,j}, \ldots, p_{N_j^v,j}\right)$$

The three nesting scenarios considered in this paper were generated by the parameters shown in **Table 4**

| Parameters | Nested | Part Crossed | More Crossed |
|---|---|---|---|
| $m_v, M_v$ | 1 , 10 | 1 , 10 | 1 , 10 |
| $\mu_v, \phi_v$ | 7.74 , 0.575 | 7.74 , 0.575 | 7.74 , 0.575 |
| $m_f, M_f$ | 0.0 , 1.1 | 0, 0.1 + $N_{HCF}$ | 0, 0.1 + $N_{HCF}$ |
| $\lambda_f$ | 0.25 | 1.0 | 2.25 ($N_{IP} = 100$) or 25.25 ($N_{IP} > 100$) |
| $\mu_\alpha, \sigma_\alpha$ | 3 , 0.2 | 3 , 0.2 | 3 , 0.2 |

**Table 4** Parameters used to simulate datasets with various levels of crossing.

## Simulation of Covariates and Design Matrices

Covariates for the Poisson regression models were simulated as independent draws as follows: simulated age, potassium, eGFR were drawn from truncated normal distributions:

$$Age_j \sim Normal(\mu_{age}, \sigma_{age})T(m_{age}, M_{age})$$

$$K_j \sim Normal(\mu_K, \sigma_K)T(m_K, M_K)$$

$$eGFR_j \sim Normal(\mu_{eGFR}, \sigma_{eGFR})T(m_{eGFR}, M_{eGFR})$$

The potassium (K) level was further "randomized" between repeated measures from the same individual. In particular, the K level from the lth visit of the jth individual was drawn from a normal distribution, with the same mean as $K_j$ and standard deviation that was proportional to $K_j$:

$$K_{l,j} \sim Normal(K_j, p_k K_j)$$

Gender and the Charlson Comorbidity Index (CCI), were drawn from a binomial and a truncated negative binomial distribution respectively:

$$Gender_j \sim Binomial(1; p_{gender})$$

$$CCI_j \sim NB(\mu_{CCI}, \sigma_{CCI})T(m_{CCI}, M_{CCI})$$

The Length of Stay of the lth visit of the jth patient was drawn from a log normal distribution:

$$LOS_{l,j} \sim Lognormal(\mu_{LOS}, \sigma_{LOS})$$

The use of bounds during covariate simulations allowed the user to limit covariate values to realistic ranges.

The parameters used to generate the fixed design matrices for the simulations are shown in **Table 5**:

| Parameters | Value |
|---|---|
| $\mu_{age}, \sigma_{age}$ | 58, 12 |
| $m_{age}, M_{age}$ | 18, 100 |
| $\mu_K, \sigma_K$ | 4.1 , 1 |
| $m_K, M_K$ | 2 , 8 |
| $p_k$ | 0.05 |
| $\mu_{eGFR}, \sigma_{eGFR}$ | 82 , 28 |
| $m_{eGFR}, M_{eGFR}$ | 15 , 120 |
| $p_{gender}$ | 0.44 |
| $\mu_{CCI}, \sigma_{CCI}$ | 0.98 , 0.55 |
| $m_{CCI}, M_{CCI}$ | 0 , 29 |
| $\mu_{LOS}, \sigma_{LOS}$ | -0.1483469, 1.413642 |

**Table 5** *Parameters used to simulate fixed effect covariate values.*

After simulating eGFR, potassium value and age for each patient, these covariates entered the model as natural splines with the boundary and interior knots shown in :

| Covariate | Boundary Knots | Interior knots |
|---|---|---|
| Age | 18, 100 | 50 , 66 |
| Potassium | 2 , 8 | 3 , 5 |
| eGFR | 15 , 120 | 50 , 90 |

**Table 6** *Position of knots used for covariates that entered the simulations through splines.*

### Simulation of Sparse Poisson Outcomes

Poisson outcomes were simulated from the design matrix of the fixed and random effects in a hierarchical fashion:

1. Simulate datasets with different number of individuals and HCF (either $N_{IP} = 100$ and $N_{HCF} = 5$ or $N_{IP} = 1000$ and $N_{HCF} = 50$) using the parameters shown in **Table 4**. Two hundred datasets were generated for the six combinations of dataset size (number of individuals and HCF) and degree of crossing (a total of 1,200 simulated datasets).
2. Simulate fixed effects covariates using the parameters and spline knot positions in **Table 5** and **Table 6** respectively.
3. The contribution of the fixed effects to the linear predictor was calculated by multiplying the matrix of the fixed effects with the vector of the fixed effects coefficients. The fixed effect coefficients used for all three scenarios are listed below (covariates entering the model as splines have more than one coefficient associated with them:
$\boldsymbol{\beta}_{age} = \{1.0, 2.0, 1.5\}$, $\boldsymbol{\beta}_K = \{1.5, -1.1, 2.2\}$, $\boldsymbol{\beta}_{eGFR} = \{1.0, 0.2, 0.12\}$, $\beta_{CCI} = 0.15$, $\beta_{Gender} = 0.26$. The coefficient associated with the intercept was used to scale the number of counts with the size of the dataset and varied according to the number of HCF and IP:
$\beta^0_{N_{HCF}=5, N_{IP}=100} = -4.5$, $\beta^0_{N_{HCF}=5, N_{IP}=1000} = -5.5$

4. The contribution of the random effects to the linear predictor was derived by cross classifying observations according to HCF and IP. For $j = 1, \ldots, N_{IP}$ and $i = 1, \ldots, N_{HCF}$ we simulated independent random effects: $u_i^{HCF} \sim Normal(0, \sigma_{HCF})$ and $u_j^{IP} \sim Normal(0, \sigma_{IP})$. These terms were added to the corresponding fixed effect term and then exponentiated to compute a rate parameter for each observation. The standard deviation of the random effects was set to $\sigma_{HCF} = 0.5$ and $\sigma_{IP} = 1.0$.
5. The rate parameter computed in the previous step was then used to draw a Poisson distribution that was truncated to lie in the interval 0 to 1.

### Simulation of Binary Observations

The hierarchical simulation for the binary outcomes proceeded as:

1. Simulate nested and crossed datasets using the parameters shown in **Table 6** for three different dataset sizes : $N_{IP} = 100$ and $N_{HCF} = 5$ or $N_{IP} = 1000$ and $N_{HCF} = 50$ or $N_{IP} = 10000$ and $N_{HCF} = 50$
2. Sample random effects from their respective distributions. There are two random effects : one at the IP and one at the HCF level: $u_{i(l)}^{HCF} \sim Normal(0, \sigma_{HCF})$, $u_j^{IP} \sim Normal(0, \sigma_{IP})$.
3. The logit of the outcome of the lth repeated measure of the jth individual, which occurred in the $i(l)$th HCF was computed as:

$$\text{logit}(p_{l,j}) = \beta^0_{N_{HCF}, N_{IP}} + u_j^{IP} + u_{i(l)}^{HCF}$$

4. The binary outcome was simulated as : $I_{j,j} \sim Bernoulli\left(\text{invlogit}(p_{l,j})\right)$.

We considered two different combinations of random effect standard deviations ("less variable" and "more variable") for each of the nine combinations of dataset size and crossing. The intercept values (which control the sparsity of the outcomes) and the standard deviations are shown in **Table 7**. Two hundred datasets were generated for each of the fifteen combinations of dataset size, degree of crossing, and variability considered for a total of 3,000 datasets.

| Dataset Size | Less Variable | More Variable |
|---|---|---|
| $N_{IP} = 100$<br>$N_{HCF} = 5$ | $\beta^0_{N_{HCF}=5, N_{IP}=100} = -5.5$,<br>$\sigma_{HCF} = 0.5$,<br>$\sigma_{IP} = 1.0$ | $\beta^0_{N_{HCF}=5, N_{IP}=100} = -5.5$,<br>$\sigma_{HCF} = 0.5$,<br>$\sigma_{IP} = 2.5$ |
| $N_{IP} = 1000$<br>$N_{HCF} = 50$ | $\beta^0_{N_{HCF}=5, N_{IP}=1000} = -7.5$<br>, $\sigma_{HCF} = 0.5$<br>, $\sigma_{IP} = 1.0$ | $\beta^0_{N_{HCF}=5, N_{IP}=1000} = -7.5$,<br>$\sigma_{HCF} = 0.5$,<br>$\sigma_{IP} = 2.5$ |
| $N_{IP} = 10000$<br>$N_{HCF} = 50$ | Not considered | $\beta^0_{N_{HCF}=5, N_{IP}=10000} = -9.5$,<br>$\sigma_{HCF} = 0.5$,<br>$\sigma_{IP} = 2.5$ |

**Table 7** Parameters used to generate binary outcomes.

**Supplementary Figure 1** Distribution of repeated measures per individual patient (IP), and number of distinct health care facilities (HCF) visited by each IP in the simulated datasets for the three different nesting scenarios considered in the paper.

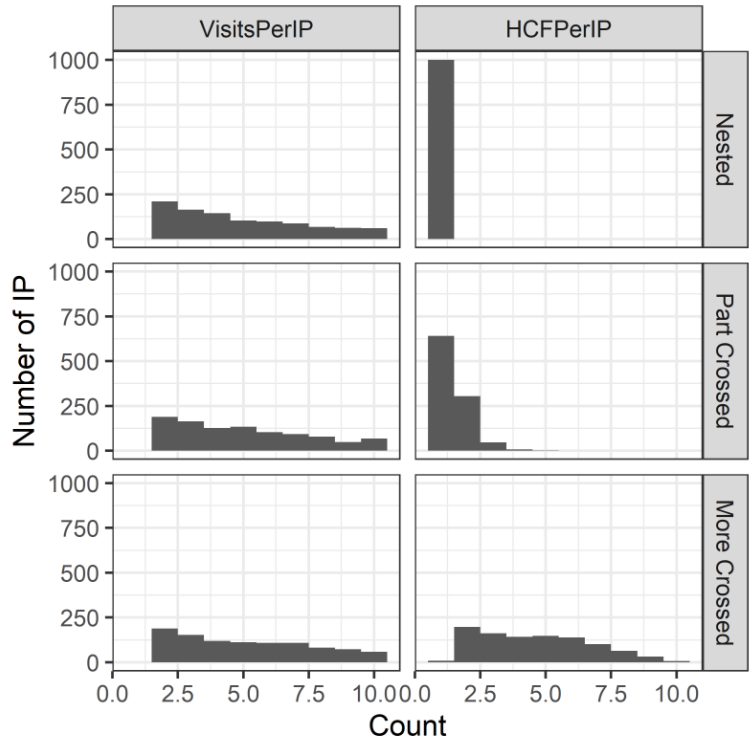

**Supplementary Figure 2** Distribution of repeated measures per health care facility (HCF) and individual patient (IP) within each HCF in the simulated datasets for the three different nesting scenarios considered in the paper.

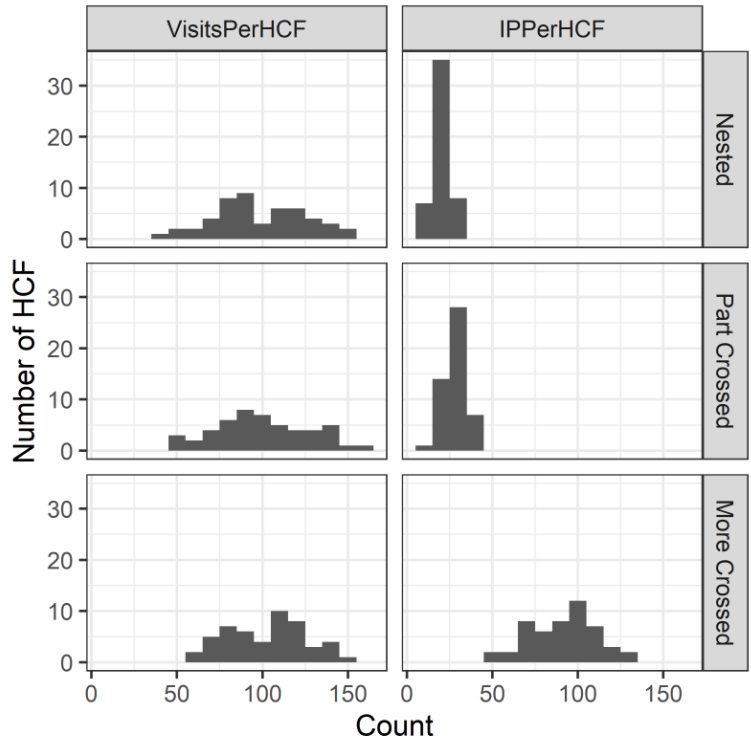

**Supplementary Figure 3** Event rates in the simulated Poisson, "less" and "more" variable binary (binomial) datasets according to nesting scenario and number of individual patients

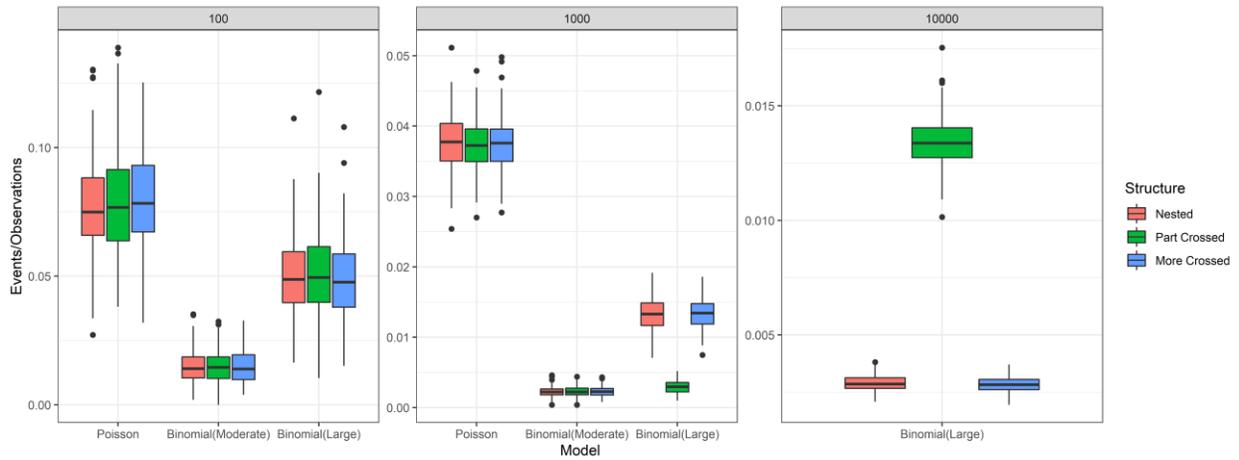

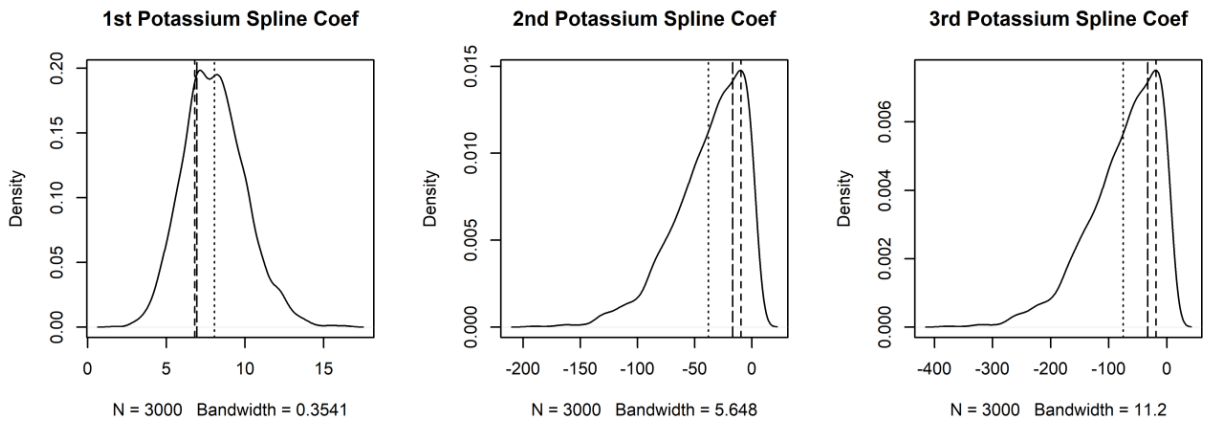

**Supplementary Figure 4**. Non-parametric kernel density estimates of the posterior marginal density of the three spline coefficients (solid curve) against the MCMC posterior mean (dotted vertical line), the AGH9 based estimated (long dash) and the h-lik estimate (dashed vertical line).